\documentclass[12pt]{article}
\usepackage{natbib}
\usepackage{url} 
\usepackage{amsmath}
\usepackage{natbib}
\usepackage[colorlinks=true,allcolors=blue]{hyperref}
\usepackage{url}
\usepackage{colonequals}

\usepackage{graphicx}
\usepackage{amsmath, amssymb}
\usepackage{bbold}
\usepackage{mathbbol}
\usepackage{algorithm}
\usepackage{algorithmic}
\usepackage{enumitem}
\usepackage{float}
\usepackage{xcolor}
\usepackage{bm} 
\usepackage{setspace}
\usepackage{multirow}
\usepackage{lscape}
\usepackage{rotating}
\usepackage{adjustbox}
\usepackage{booktabs}
\usepackage{ragged2e}
\newtheorem{theorem}{Theorem}
\usepackage[caption = false]{subfig} 
\usepackage{float}
\def\*#1{\mathbf{#1}}
\def\^#1{\amsmathbb{#1}}
\def\##1{\mathbb{#1}}
\DeclareSymbolFontAlphabet{\amsmathbb}{AMSb}%

\newcommand{\blind}{0}

\usepackage[a4paper,margin=1in]{geometry}

\begin{document}

\def\spacingset#1{\renewcommand{\baselinestretch}%
{#1}\small\normalsize} \spacingset{1}


\if0\blind
{
  \title{\bf Shape-Constrained Estimation in Functional Regression with Bernstein Polynomials}
 \author{Rahul Ghosal$^{1,\ast}$, Sujit Ghosh$^{2}$, Jacek Urbanek$^{3}$, \\Jennifer A. Schrack$^{4}$, Vadim Zipunnikov$^{5}$ \\
 $^{1}$ Department of Epidemiology and Biostatistics, University of South Carolina\\
   $^{2}$ Department of Statistics, North Carolina State University\\ 
 $^{3}$ Department of Medicine, Johns Hopkins University \\School of Medicine\\
 $^{4}$ Department of Epidemiology, Johns Hopkins Bloomberg \\School of Public Health\\
  $^{5}$ Department of Biostatistics, Johns Hopkins Bloomberg \\School of Public Health
  }
  \maketitle
} \fi

\if1\blind
{
  \bigskip
  \bigskip
  \bigskip
  \begin{center}
    {\LARGE\bf Title}
\end{center}
  \medskip
} \fi

\bigskip
\begin{abstract}
 Shape restrictions on functional regression coefficients such as  non-negativity, monotonicity, convexity or concavity are often available in the form of a prior knowledge or required to maintain a structural consistency in functional regression models. A new estimation method is developed in shape-constrained functional regression models using Bernstein polynomials. Specifically, estimation approaches from nonparametric regression are extended to functional data, properly accounting for shape-constraints in a large class of functional regression models such as scalar-on-function regression (SOFR), function-on-scalar regression (FOSR), and function-on-function regression (FOFR). Theoretical results establish the asymptotic consistency of the constrained estimators under standard regularity conditions. A projection based approach provides  point-wise asymptotic confidence intervals for the constrained estimators. A bootstrap test is developed facilitating testing of the shape constraints. Numerical analysis using simulations illustrate improvement in efficiency of the estimators from the use of the proposed method under shape constraints. Two applications include i) modeling a drug effect in a mental health study via shape-restricted FOSR and ii) modeling subject-specific quantile functions of accelerometry-estimated physical activity in the Baltimore Longitudinal Study of Aging (BLSA) as outcomes via shape-restricted quantile-function on scalar regression (QFOSR). R software implementation and illustration of the proposed estimation method and the test is provided.
\end{abstract}

\noindent%
\vfill

\newpage
\spacingset{1.5} 
\section{Introduction}
\label{sec:intro}
Functional regression \citep{Ramsay05functionaldata} is an active area of research in functional data analysis (FDA) and refers to the class of regression models with functional response and/or covariates.
Functional regression models have diverse applications in biological sciences such as genome-wide association studies (GWAS) \citep{fan2017high}, physical activity research \citep{goldsmith2016new}, functional magnetic resonance imaging \citep{reiss2017methods}, marine ecology \citep{ghosal2020variable}, radiomics \citep{yang2020quantile}, environmental modeling \citep{ghosal2021impact} and many others. Depending on whether a response or a covariate is a functional observation, functional regression models can be broadly divided into three main categories: scalar-on-function regression (SOFR), function-on-scalar regression (FOSR), and function-on-function regression (FOFR). In the simplest form of such models, the dynamic effect of the predictor of interest on the response is captured using smooth univariate or bivariate functional regression coefficients. Several  methods exist in FDA literature to estimate these regression coefficients \citep{hastie1993varying,hoover1998nonparametric,huang2004polynomial, reiss2010fast,reiss2017methods}. 

Shape restrictions such as non-negativity, monotonicity, convexity or concavity of the functional regression coefficients can either be available as a prior knowledge about the relationship between the response and the predictor of interest or be required to maintain structural consistency of such models. For example, in quantile regression analysis of systolic blood pressure (SBP) and diastolic blood pressure (DBP) \citep{kim2006quantile} on age, it is known that DBP becomes less responsive than SBP as people get older, while SBP increases. In particular, the amount of increase in DBP as a response to aging becomes progressively smaller compared to the corresponding amount of increase in SBP. Hence, for structural consistency and interpretability, the functional coefficient of DBP is required to be a nondecreasing function of the age. In the Baltimore Longitudinal Study of Aging (BLSA), the magnitude of diurnal physical activity curve was found to decrease as a function of age at all times during the day for both women and men \citep{xiao2015quantifying}. In longitudinal clinical studies exploring the effect of a drug on disease severity (e.g., \cite{ahkim2017shape}) a negative functional coefficient corresponding to the treatment group would prove the effectiveness of the drug while a negative and decreasing functional coefficient would suggest the effectiveness of the drug to increase in the follow up weeks. In a Quantile Function-on-Scalar Regression (QFOSR) framework introduced in \cite{yang2020quantile} a non-decreasing functional coefficient provide a sufficient condition \citep{yang2020random} for ensuring monotonicity of the predicted quantile functions.
In modeling of growth curves \citep{hu2009modeling}, the mean function $\mu(t)$ is required to be non-decreasing as `growth' is necessarily non-decreasing. In clinical studies, often odds-ratios of a disease can be known to be positively (negatively) associated with a functional biomarker - a knowledge that can be modelled using a constrained scalar-on-function regression (SOFR) model. Incorporation of such shape constraints on functional regression coefficients can often lead to reduced uncertainty of the coefficient estimates in the restricted parameter space \citep{lim2012consistency,yagi2020shape} and can regulate the model fit, particularly, for smaller sample sizes. Several methods have been developed for shape constrained estimation in nonparametric regression using kernel-based approaches \citep{hall2001nonparametric,dette2006simple,birke2007estimating}, smoothing splines \citep{pya2015shape}, regression splines \citep{meyer2008inference,meyer2018framework}, Bernstein polynomials  \citep{chang2005bayesian,mckay2011variable,wang2012shape} among many others. \cite{ahkim2017shape} developed a method for shape testing using constrained regression splines (B-spline) for the varying coefficient model.

In this article, we extend a Bernstein polynomial (BP) estimation approach from shape-constrained nonparametric regression \citep{wang2012shape} to a wide class of functional regression models under various shape constraints. We follow a method of sieve \citep{grenander1981abstract} and use Bernstein polynomial basis for modeling the unknown functional regression coefficients. Importantly, we show that model fitting can be reduced to solving a least square problem with linear constraints on the basis coefficients, where the constraint matrix is universal and does not depend on the order of the basis (barring dimension), observed time-points or the internal knots, unlike the constrained estimation approaches with B-splines \citep{ahkim2017shape}. This ensures the shape restrictions are satisfied everywhere over the domain and not just at the observed time points. Further, we properly account for the temporal dependence within the curves in function-on-scalar or function-on-function regression using a pre-whitening/ feasible generalized least squares approach \citep{chen2016variable,ghosal2020variable}, making the estimators more efficient. The shape constraints on the coefficient functions automatically regularizes the coefficient functions, as often required in FDA, and smoothness of the coefficient functions is achieved using a truncated basis approach  \citep{Ramsay05functionaldata,fan2015functional}, by restricting the number of BPs in the basis. A residual bootstrap based test is developed using the proposed estimation method, which can be useful for testing specific shape constraints in the absence of a prior knowledge. 

Bernstein polynomials have various attractive shape-preserving properties \citep{lorentz2013bernstein,carnicer1993shape,chang2005bayesian}. Optimal stability of BPs \citep{farouki1996optimal} makes this polynomial choice particularly suitable for modeling functional regression coefficients in the constrained functional regression problem. Theoretical results are provided on consistency of the constrained estimators under standard regularity conditions. A projection based approach is developed to construct point-wise asymptotic confidence intervals for the constrained estimators. Numerical analyses using simulations show satisfactory and competitive performance of the proposed method compared to the existing techniques for functional regression, in the presence of shape constraints. In particular, the estimates from the constrained method are shown to have reduced uncertainty in the restricted parameter space, particularly for finite sample sizes. The R code for implementation of the proposed estimation method and testing is publicly available with this article.

The rest of this article is organized as follows. We present our modeling framework, illustrate the proposed estimation method for shape constrained functional regression, establish the theoretical properties of the estimator, and propose a bootstrap test in Section \ref{sec:method1}. In Section \ref{simul}, we perform numerical simulations to evaluate the performance the proposed methods and provide comparisons with existing unconstrained functional regressions. In Section \ref{realdata}, we demonstrate application of the proposed method in two real data studies: i) a time-varying coefficient model analyzing a temporal evolution of a drug effect on the severity of illness in the National Institute of Mental Health Schizophrenia Collaborative
Study \citep{ahkim2017shape} and ii) a quantile function-on-scalar regression model of accelerometry-estimated physical activity data from the Baltimore Longitudinal Study of Aging (BLSA). We conclude in Section \ref{disc} with a brief discussion on our proposed method and some possible extensions of this work.

\section{Methodology}
\label{sec:method1}
\subsection{Modeling Framework}
We consider three types of functional regression models: scalar-on-function regression (SOFR), function-on-scalar regression (FOSR), and function-on-function regression (FOFR). Below we review these models and accompanying assumptions.

\subsubsection*{Scalar on Function Regression}
Suppose $\{Y_i,
X_{i}(t)\}$ is the observed data for the $i^{th}$ subject,
$i=1,\ldots, n$, and $Y_i$
is a scalar response of interest and
$X_i(\cdot)$ is the corresponding functional predictor. To start with, we assume the functional objects are observed on a dense and regular grid of points $S= \{t_{1},t_{2},\ldots,t_{m} \} \subset \mathcal{T}=[0,1]$, without loss of generality. Although this can be relaxed and the proposed method can be extended to accommodate more general scenarios where the functional observations are observed on an irregular and sparse domain and possibly with a measurement error. We consider the commonly used scalar-on-function regression model \citep{Ramsay05functionaldata},
\begin{eqnarray}
Y_i=\alpha+ \int_{T} X_i(t)\beta(t)dt+\epsilon_i.\label{sofr1}
\end{eqnarray}
Here, $\beta(t)$ is a smooth function over $\mathcal{T}$, capturing the dynamic effect of the functional predictor $X_i(t)$. The errors $\epsilon_i$ are assumed to be i.i.d. random variables with mean zero and variance $\sigma^2$. Note that SOFR model (\ref{sofr1}) captures only a linear effect of a single functional predictor $X_i(t)$ and multiple extensions have been proposed \citep{yao2010functional,eilers2009multivariate,mclean2014functional} to extend it to nonlinear models and models with multiple functional predictors. See \cite{reiss2017methods}, and the references therein, for a detailed review of various methods regarding the SOFR.

\subsubsection*{Function on Scalar Regression}
Let the observed data for the $i^{th}$ subject is $\{Y_i(t),X_{i}\}, i=1,\ldots, n$, where $Y_i(t)$
is now the functional response of interest and
$X_{i}$ is a corresponding scalar predictor. The commonly used function-on-scalar regression model \citep{Ramsay05functionaldata,reiss2010fast} is defined as
\begin{eqnarray}
Y_i(t)=\beta_0(t)+X_i\beta_1(t) +\epsilon_i(t).\label{FOSR1}
\end{eqnarray} 
The dependence of the functional response $Y_i(t)$ on the scalar predictor $X_i$ is captured in the function-on-scalar regression model (\ref{FOSR1}) via the coefficient function $\beta(t)$. We further assume the error functions $\epsilon_{i}(t)$ are i.i.d. copies of $\epsilon(t)$ which is a mean zero stochastic process with unknown nontrivial covariance structure. A general assumption \citep{huang2004polynomial,kim2016general} made for the error process $\epsilon(t)$ is  $\epsilon(t)=V(t) + w_t$, where $V(t)$ is a smooth mean zero stochastic process with covariance kernel $G(s,t)$ and $w_t$ is a white noise with variance $\sigma^2$. The covariance function of the error process is then given by $\Sigma(s,t)=cov\{\epsilon(s),\epsilon(t)\}=G(s,t) + \sigma^2 I(s=t)$. FOSR model (\ref{FOSR1}) assumes a linear effect of the predictor $X_i$ on $Y_i(t)$. This model has been extended to handle nonlinear associations \citep{xiao2015quantifying} and high dimensional scenarios with a focus on variable selection \citep{chen2016variable,kowal2020bayesian,ghosal2021variable}.

\subsubsection*{Function on Function Regression}
In this case, let the the observed data for the $i^{th}$ subject is $\{Y_i(t),t \in 
\mathcal{T_Y}\}$, $\{X_i(s),s \in 
\mathcal{T_X}\}, i = 1,\ldots, n$, where $Y_i(t)$ is now a functional response of interest observed over domain $\mathcal{T_Y}$ and $X_i(s)$ is a  functional predictor observed over domain $\mathcal{T_Y}$. The commonly used functional linear model (FLM) for function-on-function regression \citep{Ramsay05functionaldata,yao2005functional,wu2010varying} is defined as
\begin{eqnarray}
Y_i(t)=\beta_0(t)+\int _{\mathcal{T_X}}X_i(s)\beta_1(s,t)ds +\epsilon_i(t)\label{fof1}.
\end{eqnarray}
Here a bivariate regression coefficient  $\beta(s,t)$ captures the dependence of the functional response $Y_i(t)$ on the entire predictor trajectory $X_i(s)$, $s \in 
\mathcal{T_X}$. A special case of the the above model is when $\mathcal{T_X}=\mathcal{T_Y}$ and it is assumed the response $Y_i(t)$ depends on $X_i(\cdot)$ concurrently. Specifically, $\beta_1(s,t)=\beta_1(t) I(s=t)$. The resulting functional linear concurrent model \citep{Ramsay05functionaldata} is given by 
\begin{eqnarray}
Y_i(t)=\beta_0(t)+X_i(t)\beta_1(t) +\epsilon_i(t)\label{FLCM1}.
\end{eqnarray}
In both regression models discussed above, $\epsilon(\cdot)$ is assumed to be a mean zero stochastic process with an unknown nontrivial covariance structure. Multiple extensions have been proposed involving nonlinear associations and multiple predictors in the function-on-function regression model \citep{kim2016general,scheipl2015functional,kim2018additive}.

\subsection{Shape Constrained Functional Regression using Bernstein Polynomials}
\label{shape reg}
 In all the functional regression models discussed above, the dependence between the response and predictors are captured using nonparametric functional coefficients. Often a prior knowledge about these functional regression coefficients is available in the form of constraints such as $\beta_1(t)>0$, $\beta_1(t)$ is increasing, $\beta_1(t)$ is convex (concave), $\beta_1(s,t)$ is monotone or bi-monotone, etc. Incorporation of these constraints in the estimation procedure can lead to a reduced uncertainty about estimates in the restricted parameter space. Below, we develop a general purpose estimation procedure for the functional regression models \ref{sofr1}-\ref{FLCM1} under such shape constraints. We express any univariate coefficient functions $\beta(t)$
in models (\ref{sofr1}), (\ref{FOSR1}), (\ref{fof1}), and (\ref{FLCM1}) in terms of univariate expansions of Bernstein basis polynomials. Specifically, we model them as follows:
\begin{equation}
    \beta(t)=\sum_{k=0}^{N}\beta_{k}b_k(t,N),\hspace{2mm} \textit{where}\hspace{2mm}b_k(t,N)={N \choose k}t^k(1-t)^{N-k}, \hspace{2mm} \textit{for}\hspace{2mm } 0\leq t\leq 1.
    \label{uni}
\end{equation}
The number of basis polynomials depends on the order of the polynomial basis $N$. Note that $b_k(t,N)\geq 0$ and $\sum_{k=0}^{N}b_k(t,N)=1$. Let $\mathcal{F}$ be the class of shape restricted functions we are interested in. Following \cite{wang2012shape}, we define the constrained Bernstein polynomial sieve as follows:
\begin{equation}
    \mathcal{F}_{N}=\{B_{N}(t)=\sum_{k=0}^{N}\beta_{k}b_k(t,N): \^A_{N}\bm\beta_{N}\geq \bm 0, \sum_{k=0}^{N}|\beta_{k}|\leq L_{N}\}, \label{sieve}
\end{equation}
where $\bm\beta_{N}=(\beta_{0},\beta_{1},\ldots,\beta_{N})^T$ are the unknown basis coefficients and $\^A_{N}$ is the constraint matrix (of the dimension $R_N\times(N+1)$) chosen in a way to guarantee the desired shape restriction (i.e., $\mathcal{F}_{N}\subset \mathcal{F}$). Note that the condition  $\sum_{k=0}^{N}|\beta_{k}|\leq L_{N}$ was only required for establishing asymptotic properties implying that the functions spanned by this basis are bounded in absolute value by $L_N$, and can be avoided in practice \citep{wang2012shape}. 


Bivariate function $\beta(s,t)$ can be modelled using a bivariate basis expansion with a tensor product of univariate Bernstein polynomials as follows:
\begin{equation}
    \beta(s,t)=\sum_{k_1=0}^{N}\sum_{k_2=0}^{N}\beta_{k_1,k_2}b_{k_1}(s,N) b_{k_2}(t,N),\hspace{2mm} \textit{where}\hspace{2mm}b_{k_j}(x,N)={N \choose k_j}x^{k_j}(1-x)^{N-k_j},\label{bibern}
\end{equation}
where $s,t \in [0,1]$. In this case, we can define the sieve $\mathcal{F}_{N}$ as
\begin{equation}
    \mathcal{F}_{N}=\{B_{N}(s,t)=\sum_{k_1=0}^{N}\sum_{k_2=0}^{N}\beta_{k_1,k_2}b_{k_1}(s,N) b_{k_2}(t,N): \^A_{N}\bm\beta_{N}\geq \bm 0, \sum_{k_1=0}^{N}\sum_{k_2=0}^{N}|\beta_{k_1,k_2}|\leq L_{N}\}. \label{sievebi}
\end{equation}
Here  $\bm\beta_{N}$ denotes the stacked vector $\{\beta_{k_1,k_2}\}_{k_1=0,k_2=0}^{N,N}$ and $\^A_{N}$ is the constraint matrix of dimension $R_N\times(N+1)^2$ ensuring the required shape restriction on the surface $\beta(s,t)$.

\hspace*{-8 mm}
\textit{Remark 1:}\\
For notational simplicity, we denote the order of Bernstein polynomial by $N$ in both the variables $s,t$. Below, we consider the most common scenarios for constraints on $\beta(t)$ and $\beta(s,t)$ defined by $\^A_{N}$, including:   nonnegativity, monotonicity, convexity/concavity and their combinations.

\subsubsection*{Properties of Bernstein polynomial sieve}
The sequence of function spaces $\mathcal{F}_{N}$ is nested in $\mathcal{F}$
and $\bigcup_{N=1}^{\infty}\mathcal{F}_{N}$ is dense in $\mathcal{F}$ with respect to the sup-norm (see property 3.1 and 3.2 in \cite{wang2012shape}). This result along with the Stone-Weierstrass approximation theorem guarantee that for any $\beta(t)\in \mathcal{F}$, there exists $B_N(t)\in \mathcal{F}_{N} \subset \bigcup_{j=1}^{\infty}\mathcal{F}_{j}$ which converges uniformly \citep{lorentz2013bernstein} to $\beta(t)$.

\subsubsection*{Constraints}
\begin{itemize}
    \item \underline{Fixed boundaries}\\
    Let $\beta(t)$ be in the space $\mathcal{F}=\{\beta\in C[0,1]:\beta(0)=a_0, \beta(1)=a_1\}$, where $C[0,1]$ is the class of all continuous functions on $[0,1]$. For any $B_N(t)$ in the corresponding sieve $\mathcal{F}_{N}$ (\ref{sieve}), these boundary conditions reduce to linear equality constraints of the form 
     \begin{equation*}
 \^A_{N}\bm\beta_{N}\equiv \begin{pmatrix}
 1 & 0 & \dots & 0\\
 0 & 0 & \dots &1\\
  \end{pmatrix} _{2\times (N+1)} \begin{pmatrix}
 \beta_{0} \\
 \beta_{1}\\
 \vdots\\
 \beta_{N} 
 \end{pmatrix} =  \begin{pmatrix}
 a_0 \\
 a_1\\
 \end{pmatrix}.
 \end{equation*}
Thus, $\beta_0 = a_0$ and $\beta_N = a_1$. Here  $\^A_{N}$ is the constraint matrix with rank $R_N=2$. Note that this equality constraint can be decomposed into combination of two inequality contraints in the usual way, i.e., $\^A_{N}\bm\beta_{N} \ge \bm a$ and $-\^A_{N}\bm\beta_{N} \ge -\bm a$, where $\bm a = (a_0, a_1)$.
\item \underline{Nonnegativity}\\
    Let $\beta(t)$ be a nonnegative function in the space $\mathcal{F}=\{\beta\in C[0,1]:\beta(t)\geq0 \hspace{3 mm} \forall t \in [0,1]\}$, where $C[0,1]$ is the class of all continuous functions on $[0,1]$. For any $B_N(t)$ in the corresponding sieve $ \mathcal{F}_{N}$ (\ref{sieve}), the nonegativity constraint reduces to the linear inequality constraint
    
  \begin{equation*}
 \^A_{N}\bm\beta_{N}\equiv \begin{pmatrix}
 1 & 0 & \dots & 0\\
 0 & 1 & \dots &0\\
 & & \ddots & &\\
 0 &0& \dots & 1\\
 \end{pmatrix} _{(N+1)\times (N+1)} \begin{pmatrix}
 \beta_{0} \\
 \beta_{1}\\
 \vdots\\
 \beta_{N} 
 \end{pmatrix} \geq  \begin{pmatrix}
 0 \\
 0\\
 \vdots\\
0
 \end{pmatrix}.
    \end{equation*}
Here  $\^A_{N}$ is the constraint matrix with rank $R_N=(N+1)$.   

\item \underline{Monotonicity}\\  
  Let $\beta(t)$ be a monotone (non-decreasing) function in the space $\mathcal{F}=\{\beta\in C[0,1]:\beta(t_1)\leq \beta(t_2) \hspace{3 mm} \forall 0\leq t_1 \leq t_2 \leq 1 \}$. Note that for any $B_N(t)$ in the corresponding sieve $ \mathcal{F}_{N}$ (\ref{sieve}), its derivative is given by $B_{N}^{\prime}(t)=N\sum_{k=0}^{N-1}(\beta_{k+1}-\beta_{k})b_k(t,N-1)$. Hence if $\beta_{k+1}\geq\beta_{k}$ for $k=0,1,\ldots, N-1$,   $B_{N}(t)$ is non decreasing and $\mathcal{F}_{N}\subset \mathcal{F}$. Thus the linear constraint on the parameters is given by,
  \begin{equation*}
 \^A_{N}\bm\beta_{N}\equiv \begin{pmatrix}
 -1 & 1&0  & \dots & 0\\
 0 & -1 &1 &0 & \dots\\
 & & \ddots & &\\
 0 &\dots &0& -1 & 1\\
 \end{pmatrix} _{N\times (N+1)} \begin{pmatrix}
 \beta_{0} \\
 \beta_{1}\\
 \vdots\\
 \beta_{N} 
 \end{pmatrix} \geq  \begin{pmatrix}
 0 \\
 0\\
 \vdots\\
0
 \end{pmatrix}.
    \end{equation*}
 Here  $\^A_{N}$ is the constraint matrix with rank $R_N=N$. Note that a non-increasing constraint on $\beta(t)$ can simply be obtained by reversing the inequality. 
 
 \item \underline{Convexity/Concavity}\\ 
Let $\beta(t)$ be a convex function in the space $\mathcal{F}=\{\beta\in C[0,1]:2\beta(\frac{t_1+t_2}{2})\leq \beta(t_1)+\beta(t_2), \hspace{3 mm} \forall t_1, t_2 \in [0,1] \}$. Note that for any $B_N(t)$  in the sieve the second derivative is given by $B_{N}^{\prime}(t)=N(N-1)\sum_{k=0}^{N-2}(\beta_{k+2}-2\beta_{k+1}+\beta_k)b_k(t,N-2)$. Hence if $\beta_{k+2}-2\beta_{k+1}+\beta_k \geq 0$ for $k=0,1,\ldots, N-2$,   $B_{N}^{\prime\prime}(t)\geq 0$  and $\mathcal{F}_{N}\subset \mathcal{F}$. Hence the convexity constraint on the coefficient function reduces to the following linear inequality constraint,  where  $\^A_{N}$ is the constraint matrix with rank $R_N=N-1$. A concave constraint on $\beta(t)$ can simply be obtained by reversing the inequality.
 \begin{equation*}
 \^A_{N}\bm\beta_{N}\equiv \begin{pmatrix}
 1 & -2&1  & \dots & 0\\
 0 & 1 &-2 &1 & \dots\\
 & & \ddots & &\\
 0 &\dots &1& -2 & 1\\
 \end{pmatrix} _{(N-1)\times (N+1)} \begin{pmatrix}
 \beta_{0} \\
 \beta_{1}\\
 \vdots\\
 \beta_{N} 
 \end{pmatrix} \geq  \begin{pmatrix}
 0 \\
 0\\
 \vdots\\
0
 \end{pmatrix}.
\end{equation*}

 \item \underline{Bivariate monotonicity}\\ 
Let $\beta(s,t)$ be a bivariate function monotone in both coordinates, specifically, $\mathcal{F}=\{\beta\in C[0,1]^2:\beta(s_1,t_1)\leq \beta(s_2,t_1), \beta(s_1,t_1)\leq \beta(s_1,t_2),  \hspace{3 mm} \forall 0\leq s_1 \leq s_2 \leq 1, 0\leq t_1 \leq t_2 \leq 1 \}$. Here, $C[0,1]^2$ is the class of all continuous functions on $[0,1]^2$. For any $B_N(s,t)$ in the sieve $ \mathcal{F}_{N}$ (\ref{sievebi}), the partial derivatives are given by $\frac{\partial B_N}{\partial s}=N\sum_{k_1=0}^{N-1}\sum_{k_2=0}^{N}(\beta_{k_1+1,k_2}-\beta_{k_1,k_2})b_{k_1}(s,N-1) b_{k_2}(t,N)$ and $\frac{\partial B_N}{\partial t}=N\sum_{k_1=0}^{N}\sum_{k_2=0}^{N-1}\\(\beta_{k_1,k_2+1}-\beta_{k_1,k_2})b_{k_1}(s,N) b_{k_2}(t,N-1)$. Hence the bimonotone constraint redcues to a linear constraint of the form, $\^A_{N}\bm\beta_{N}\geq \bm 0$, where the constraint matrix is given by $\^A_{N}=\begin{pmatrix}
 \^A_{N}^{(1)} \\
 \^A_{N}^{(2)}\\
 \end{pmatrix}$. The first submatrix $\^A_{N}^{(1)}$ ensures monotonicity in $s$ and $\^A_{N}^{(2)}$ ensures monotonicity in $t$. The two submatrices are given by
  \begin{equation*}
\^A_{N}^{(1)} =\begin{pmatrix}
 -1 & 0& \dots &0 &1  & & & & & \\
  & -1 &0 &\dots &0 &1 & & & &  \\
 & & & & \ddots & &\\
 & & & &-1 &0 & \dots & 0 & 1\\
 \end{pmatrix} _{N(N+1)\times (N+1)^2} ,
    \end{equation*}
 and    
 \begin{equation*}
\^A_{N}^{(2)} =\begin{pmatrix}
 \^B & & & & & \\
  & &\^B & & & &  \\
 & & & \ddots & &\\
 & & & & &\^B \\
 \end{pmatrix} _{N(N+1)\times (N+1)^2} \^B= \begin{pmatrix}
 -1 & 1&0  & \dots & 0\\
 0 & -1 &1 &0 & \dots\\
 & & \ddots & &\\
 0 &\dots &0& -1 & 1\\
 \end{pmatrix} _{N\times (N+1)} 
    \end{equation*}
respectively. If monotonicity is required only in one of the coordinate $s$ or $t$, then the constraint matrix can be taken to be  $\^A_{N}=\^A_{N}^{(1)}$ or $\^A_{N}=\^A_{N}^{(2)}$ accordingly.

\item \underline{Partial convexity of $\beta(s.t)$}\\ 
Suppose $\beta(s,t)$ is a convex function in $s$ for every fixed $t$ and vice-versa. Here the restricted function space is given by 
$\mathcal{F}=\{\beta\in C[0,1]:2\beta(\frac{s_1+s_2}{2},t1)\leq \beta(s_1,t_1)+\beta(s_2,t_1)$ and $\hspace{2 mm} 2\beta(s_1,\frac{t_1+t_2}{2})\leq \beta(s_1,t_1)+\beta(s_1,t_2) \hspace{1 mm} \forall s_!,s_2,t_1, t_2 \in [0,1]  \}$. Note that, for any $B_N(s,t)$ in the sieve $ \mathcal{F}_{N}$ (\ref{sievebi}), the partial derivatives are given by $\frac{\partial^2 B_N}{\partial s^2}=N\sum_{k_1=0}^{N-2}\sum_{k_2=0}^{N}(\beta_{k_1+2,k_2}-2\beta_{k_1+1,k_2}+\beta_{k_1,k_2})b_{k_1}(s,N-2) b_{k_2}(t,N)$ and $\frac{\partial^2 B_N}{\partial t^2}=N\sum_{k_1=0}^{N}\sum_{k_2=0}^{N-2} (\beta_{k_1,k_2+2}-2\beta_{k_1,k_2+1}+\beta_{k_1,k_2})b_{k_1}(s,N) b_{k_2}(t,N-2)$. Hence the partial convexity constraints reduced to linear constraints of the form, $\^A_{N}\bm\beta_{N}\geq \bm 0$, where the constraint matrix is given by $\^A_{N}=\begin{pmatrix}
 \^A_{N}^{(1)} \\
 \^A_{N}^{(2)}\\
 \end{pmatrix}$. The first submatrix $\^A_{N}^{(1)}$ ensures convexity in $s$ and $\^A_{N}^{(2)}$ ensures convexity in $t$.
 The two submatrices are given by
  \begin{equation*}
\^A_{N}^{(1)} =\begin{pmatrix}
 1 & 0& \dots 0 &-2 &0\dots &0 &1  \\
  & 1 &0 &\dots 0 &-2 &0\dots &0 &1\\
 & & & & \ddots & &\\
 \end{pmatrix} _{(N^2-1)\times (N+1)^2} ,
    \end{equation*}
 and    
 \begin{equation*}
\^A_{N}^{(2)} =\begin{pmatrix}
 \^B & & & & & \\
  & &\^B & & & &  \\
 & & & \ddots & &\\
 & & & & &\^B \\
 \end{pmatrix} _{(N^2-1)\times (N+1)^2} \^B=\begin{pmatrix}
 1 & -2&1  & \dots & 0\\
 0 & 1 &-2 &1 & \dots\\
 & & \ddots & &\\
 0 &\dots &1& -2 & 1\\
 \end{pmatrix} _{(N-1)\times (N+1)}
    \end{equation*}
respectively.

\item Various other shape constraints on $\beta(t)$ and $\beta(s,t)$ including any combination of above constraints can similarly be shown to be reduced to linear inequality constraints of the form $\^A_{N}\bm\beta_{N}\geq \bm 0$.
\end{itemize}

\subsubsection*{Scalar response regression}
Using the basis expansion for the coefficient function $\beta(t)$ in (\ref{uni}), the SOFR model (\ref{sofr1}) can be reformulated as follows
\begin{align}
    Y_i&=\alpha+ \int_{T} X_i(t)\beta(t)dt+\epsilon_i \notag\\
    &=\alpha+ \sum_{k=0}^{N}\beta_{k} \int_{T} X_i(t) b_k(t,N)dt+\epsilon_i \notag\\
    &=\alpha+ \sum_{k=0}^{N}\beta_{k} W_{ik}+\epsilon_i, \hspace{2mm} \textit{where $ W_{ik}=\int_{T} X_i(t) b_k(t,N)dt$ }\notag\\
   &=\alpha+\*W_i^T\bm\beta+\epsilon_i, \hspace{2mm} \label{sofrrep1}
\end{align}
where $ \*W_{i}=(W_{i0},W_{i1},\ldots,W_{iN})^T$ and $\bm\beta=(\beta_{0},\beta_{1},\ldots,\beta_{N})^T$. Parameters $(\alpha,\bm\beta)$ is estimated by minimizing the constrained least square problem,
\begin{equation}
    (\hat{\alpha},\hat{\bm\beta})=\underset{\alpha,\bm\beta}{\text{argmin}}\hspace{2 mm}  \sum_{i=1}^{n}({Y_i}-\alpha-\*W_i^T\bm\beta)^{2} \hspace{4mm} \textit{s.t \hspace{ 4 mm} $\^A_{N}\bm\beta\geq \bm 0$}, \label{sofroptorig}
\end{equation}
where the constraint $\^A_{N}\bm\beta\geq \bm 0$ corresponds to the required shape restriction on $\beta(t)$. The above optimization problem is a quadratic programming problem \citep{goldfarb1982dual,goldfarb1983numerically} and can be efficiently solved in R using the \texttt{quadprog} \citep{qp} or the \texttt {restriktor} \citep{res} package. Additional scalar covariates of interest $\*Z_i$ (confounders) can be readily included in the SOFR model (\ref{sofr1}) and the above optimization criterion
through an additional term $\*Z_i^T\bm\gamma$ \citep{reiss2017methods} capturing effects of the scalar predictors.

\subsubsection*{Functional response regression}
We use the Bernstein polynomial basis expansions for modeling univariate and bivariate coefficient functions $\beta_0(t)$, $\beta_1(t)$ and $\beta_1(s,t)$ in function-on-scalar regression model (\ref{FOSR1}) and function-on-function regression models (\ref{fof1}),(\ref{FLCM1}).  We denote the stacked functional response corresponding to subject $i$ as $\*Y_i=(Y_i(t_1),Y_i(t_2),\ldots, Y_i(t_m))$. Using the basis expansions for the coefficient functions, the function-on-scalar or the function-on-function regression models can be reformulated as follows
\begin{equation}
    \*Y_i= \^B_0\bm\beta_0+\^W_i\bm\beta_1 + \bm\epsilon_i. \label{func response}
\end{equation}
Here, the matrix $\^B_0$ depends on the basis functions used for modeling $\beta_0(t)$. Matrix $\^W_i$ depends on the basis functions used for $\beta_1(t)$ or $\beta_1(s,t)$ and the corresponding predictor $X_i$, $X_i(t)$, or the entire trajectory $X_i(\cdot)$. Vectors $\bm\beta_0$ and $\bm\beta_1$ denote the basis coefficients, and the stacked residuals are denoted as $\bm\epsilon_i=(\epsilon_i(t_1),\epsilon_i(t_2),\ldots, \epsilon_i(t_m))$. The shape restrictions on the coefficient function of interest $\beta_1(t)$, or $\beta_1(s,t)$ can be specified as linear constraints of the form $\^A\bm\beta_1\geq 0$ as illustrated in the Section \ref{shape reg}. As mentioned earlier, the error process $\epsilon(\cdot)$ is assumed to have a nontrivial covariance kernel $\Sigma(s,t)$ for the functional regression models with a functional response. To take into account the within curve dependence while doing estimation, we propose the following two-step method .

\hspace* {- 8 mm}
\textbf{Step 1}\\
Note that the covariance function of the error process is given by $\Sigma(s,t)=cov\{\epsilon(s),\epsilon(t)\}\\=G(s,t) + \sigma^2 I(s=t)$. For data observed on dense and regular grid, the covariance matrix of the residual vector $\bm\epsilon_i$ is $\#\Sigma_{m\times m}$, the covariance kernel $\Sigma(s,t)$ evaluated on the grid $S= \{t_{1}, t_{2},\ldots, t_{m} \}$. In reality $\#\Sigma_{m\times m}$ is unknown, and we need an estimator $\hat{\#\Sigma}_{m\times m}$. In the context of functional data, we want to estimate $\Sigma(\cdot,\cdot)$ nonparametrically.
If the original residuals $\epsilon_{ij}$ were available, functional principal component analysis (FPCA) can be used, e.g., \cite{yao2005functional2} to estimate $\Sigma(s,t)$. By Mercer's theorem, the covariance kernel $G(s,t)$ has a spectral decomposition
$$G(s,t)=\sum_{k=1}^{\infty}\lambda_k\phi_k(s)\phi_k(t),$$
where $\lambda_1\geq\lambda_2\geq \ldots0$ are the ordered eigenvalues and $\phi_k(\cdot)$s are the corresponding eigenfunctions. Thus we have the decomposition $\Sigma(s,t)=\sum_{k=1}^{\infty}\lambda_k\phi_k(s)\phi_k(t) + \sigma^2 I(s=t)$. Given $\epsilon_{t_{ij}}=V(t_{ij}) + w_{ij}$, FPCA \citep{yao2005functional} can be used to get $\hat{\phi}_k(\cdot)$, $\hat{\lambda}_k$s and $\hat{\sigma}^2$. So an estimator of $\Sigma(s,t)$ can be formed as
$$\hat{\Sigma}(s,t)=\sum_{k=1}^{K}\hat{\lambda}_k\hat{\phi}_k(s)\hat{\phi}_k(t) + \hat{\sigma}^2 I(s=t),$$ 
where $K$ is large enough such that percent of variance explained (PVE) by the selected eigencomponents exceeds some pre-specified value such as $99\%$ or $95\%$. 

In practice, we don't have  the original residuals $\epsilon_{ij}$.
Hence we fit an unconstrained model by minimizing the residual sum of squares $\sum_{i=1}^{n}||{\*Y_i}-\^B_0\bm\beta_0-\^W_i\bm\beta_1||_2^{2}$,
and obtain the residuals $e_{ij}=Y_{i}(t_j)-\hat{Y_{i}}(t_j)$. Then treating $e_{ij}$ as our original residuals, we obtain $\hat{\Sigma}(s,t)$ and $\hat{\#\Sigma}_{m\times m}$ using the FPCA approach describe above. 

\hspace* {- 8 mm}
\textbf{Step 2}\\
We pre-whiten \citep{chen2016variable} $\*Y_i$, $\^B_0$ and $\^W_i$ in model (\ref{func response}) using the estimated covariance matrix $\hat{\#\Sigma}^{-1/2}_{m\times m}$ as $\*Y_i^*=\hat{\#\Sigma}^{-1/2}_{m\times m} \*Y_i$, $\^B_0^*=\hat{\#\Sigma}^{-1/2}_{m\times m} \^B_0$,  $\^W_i^*=\hat{\#\Sigma}^{-1/2}_{m\times m} \^W_i$. Subsequently, the model parameters are estimated from the constrained optimization problem

\begin{equation}
     (\hat{\bm\beta_0},\hat{\bm\beta_1})=\underset{\bm\beta_0,\bm\beta_1}{\text{argmin}}\hspace{2 mm}  \sum_{i=1}^{n}||{\*Y_i^*}-\^B_0^*\bm\beta_0-\^W_i^*\bm\beta_1 ||_2^{2} \hspace{4mm} \textit{s.t \hspace{ 4 mm} $\^A\bm\beta_1\geq \bm 0$}. \label{funopt}
\end{equation}
Again the above constrained least square optimization can be performed using quadratic programming as in the case of shape constrained scalar-on-function regression. 

\hspace*{-8 mm}
\textit{Remark 2:}\\
This feasible GLS approach is used in the article for all the results corresponding to functional response regression. The constrained estimator can be viewed as a projection of the unconstrained GLS estimator as illustrated in Appendix B of the Supplementary Material.

\subsubsection*{Consistency of the shape constrained estimators}
We establish the consistency of shape constrained estimator in scalar-on-function regression. The functional response model is considered in Appendix B of the Supplementary Material.

\begin{theorem}
Consider scalar-on-function regression model (\ref{sofr1}). Suppose the following conditions hold.
\begin{enumerate}[label=(H\arabic*)]
    \item $||X(\cdot)||_{\mathcal{L}_2}\leq C_1< \infty,$ a.s.
    \item $Var(Y\mid X(\cdot)=x(\cdot)) \leq C_2 < \infty$ a.s.
    \item The eigenvalues of the covariance operator of $X(\cdot)$ are positive and distinct.
    \item The functional coefficient $\beta(t)$ defined on $[0, 1]$ is supposed to be sufficiently smooth. In particular, let $\mathcal{H}$ be the class of functions $\beta(\cdot)$ having $q\geq 0$ derivatives, with $\beta^{(q)}(t)$ satisfying $|\beta^{(q)}(t_1)-\beta^{(q)}(t_2)|\leq C_3|t_1-t_2|^v$, $C_3>0$ and $v\in [0,1]$.
   
    \item$\lim_{n}d(\beta(\cdot),\mathcal{F}_{N})=0$ and $\lim_{n}Nlog N/n=0$, where $d(\beta(\cdot),\mathcal{F}_{N})$ is defined as $d(\beta(\cdot),\mathcal{F}_{N})=\inf_{g\in \mathcal{F}_{N}} \sup_{t\in \mathcal{T}} |\beta(t)-g(t)|$. 
    \end{enumerate}
If the shape restriction assumption for  $\beta(t)$ holds, i.e., the true coefficient function $\beta(t) \in \mathcal{F}\cap\mathcal{H}$, then the constrained estimator $\hat{\beta}_{c}(t)$ is a consistent estimator of $\beta(t)$.
\label{thm:eqvalance}
\end{theorem}

\hspace*{- 8 mm}
\textbf{Proof:}
The proof of Theorem 1 is given in Appendix A of the Supplementary Material.

\hspace*{-8 mm}
\textit{Remark 3:}\\
The primary advantage of the proposed estimation method, particularly for finite sample sizes, comes from the potential reduction in variance of the
constrained estimator, as the objective function is minimized over a constrained (smaller) space with a lower entropy. This point is also well illustrated in our empirical analysis.

\subsection{Uncertainty Quantification}
As shown in Section 2.2, the constrained estimator can be viewed as the projection of the unconstrained estimator onto the restricted space: $\hat{\bm\beta}_{r}=\underset{\bm\beta \in \bm\Theta_R }{\text{argmin}}\hspace{2 mm} ||\bm\beta-\hat{\bm\beta}_{ur}||^{2}_{\hat{\*\Omega}}.$ Hence, we can use the projection of the large sample distribution of $\sqrt{n}(\hat{\bm\beta}_{ur}-\bm\beta^0)$ to approximate the distribution of 
$\sqrt{n}(\hat{\bm\beta}_{r}-\bm\beta^0)$. We assume that $\sqrt{n}(\hat{\bm\beta}_{ur}-\bm\beta^0)$ is asymptotically distributed as $N(0,\bm\Delta)$ under suitable regularity conditions (analogous to assumption 2 of Theorem 1 in \cite{freyberger2018inference}), where $\bm\Delta$ can be estimated by a consistent estimator. 
For example $\hat{\bm\Delta}= (\hat{\bm\Omega})^{-1}(\frac{1}{n}\sum_{i=1}^{n}\hat{\epsilon_i}^2\*W_i\*W_i^T)(\hat{\bm\Omega})^{-1}$ for the scalar response case. Let the Bernstein polynomial approximation of $\beta(t)$ be given by $\beta_N(t)=\sum_{k=0}^{N}\beta_{k}b_k(t,N)=\rho_{K_n}(t)^{'}\bm\beta$. Algorithm \ref{algo 12} will be used to obtain a point-wise approximate $100(1-\alpha)\%$ asymptotic confidence interval for the true coefficient function $\beta^0(t)$.

\begin{algorithm}[ht]
\label{algo2}
\caption{Point-wise confidence interval of  $\beta^0(t)$ under shape restriction}
\begin{algorithmic}
\label{algo 12}
\STATE 1. Fit the unconstrained model and obtain the unconstrained estimator $\hat{\bm\beta}_{ur}=\underset{\bm\beta \in R^{K_n} }{\text{argmin}}\hspace{2 mm}  \sum_{i=1}^{n}({Y_i}-\*W_i^T\bm\beta)^{2}$ (for scalar response) or $\hat{\bm\beta}_{ur}=\underset{\bm\beta \in R^{K_n} }{\text{argmin}}\hspace{2 mm}  \sum_{i=1}^{n}||{\*Y_i^*}-\^Z_i^*\bm\beta ||_2^{2}$ (for the functional response, $\^Z_i^*=[\^B_0^* \hspace{2mm}\^W_i^*] $). 

\STATE 2. Let the estimated asymptotic covariance matrix of the unconstrained estimator be given by $\hat{\bm\Delta}_n=\hat{\bm\Delta}/n= \hat{cov}(\hat{\bm\beta}_{ur})$. 
\STATE 3. For $b = 1$ to $B$ 
\STATE [-]  generate $\*Z_b\sim N_{K_n}(\hat{\bm\beta}_{ur},\hat{\bm\Delta}_n)$.
\STATE [-] compute the projection of $\*Z_b$ as $\hat{\bm\beta}_{r,b}=\underset{\bm\beta \in \bm\Theta_R }{\text{argmin}}\hspace{2 mm} ||\bm\beta-\*Z_b||^{2}_{\hat{\*\Omega}}.$
\STATE [-] End For
\STATE 4. $100(1-\alpha)\%$ point-wise confidence interval for $\beta^0(t)$ is given by $(C_{\alpha/2}(t), C_{1-\alpha/2}(t) )$, where $C_{\alpha}(t)$ denotes the empirical $\alpha^{th}$ percentile of $\rho_{K_n}(t)^{'}\hat{\bm\beta}_{r,b}^{j}$ ($b=1,\ldots,B$), and $j=0,1$.
\end{algorithmic}
\end{algorithm}


\subsection{Testing for Shape Constraints}
So far we have focused on estimation under a prior knowledge of shape constraints on functional regression coefficients. In many cases, such constraints may not be known beforehand or a practitioner might posit some prior beliefs about the shape, which he or she would like to test. In this section, we develop a testing procedure for shape constraints based on the proposed estimation method. In particular, we use bootstrap on a F-type test statistic based on the residual sum of squares of the constrained (null) and unconstrained (full) model similar to \cite{kim2016general}. Alternatively, one may also use the idea of wild bootstrap \citep{davidson2008wild} which generates the responses using scaled residuals only. The test statistic is defined as 
\begin{equation}T=\frac{RSS_c-RSS_u}{RSS_u}, \label{boot}\end{equation} where $RSS_c,RSS_u$ are the residual sum of squares under the constrained and unconstrained model respectively.
For the scalar response case $RSS_u=\sum_{i=1}^{n}({Y_i}-\hat\alpha_u-\*W_i^T\hat{\bm\beta}_u)^{2}$, where  $(\alpha_u,\hat{\bm\beta}_u)$ are the unconstrained estimators and $RSS_c$ is defined analogously.
For models with functional response $RSS_u$, $RSS_c$ is calculated based on the residual sum of squares from model (\ref{func response}) after obtaining unconstrained and constrained estimates of ${\bm\beta}_j$ ($j=0,1$).
We illustrate our testing procedure for models with scalar response and functional response with univariate regression functions $\beta(\cdot)$. The case with bivariate coefficient functions $\beta(\cdot,\cdot)$ for more general function-on-function regression models (e.g., FLM) can be handled similarly.

\subsubsection*{Shape Testing with Scalar Response}
We consider the SOFR model (\ref{sofr1}). We are interested in testing
prior shape restrictions on the functional coefficient  $\beta(t)$ for $t\in [0, 1]$. Let $\mathcal{F}$ be the class of shape restricted functions we are interested in and $\mathcal{H}$ be space of function as defined in condition (H4) in Theorem \ref{thm:eqvalance}. We want to test the null hypothesis
$$H_{0} : \beta(\cdot) \in \mathcal{F}\cap\mathcal{H} \hbox{ \;\; versus \;\; } H_{1} : \beta(\cdot)\in \mathcal{H}.$$ The null distribution of the the test statistic $T$ in (\ref{boot}) is approximated using bootstrap. We present the complete bootstrap procedure in Algorithm \ref{algo 1}.

\begin{algorithm}[ht]
\label{algo1}
\caption{Bootstrap algorithm for shape testing with scalar response}
\begin{algorithmic}
\label{algo 1}
\STATE 1. Fit the unconstrained SOFR model (\ref{sofr1}) using Bernstein-polynomial representation in (\ref{sofrrep1}) and calculate the residuals $e_i=Y_i-\hat{Y}_i$, for $i=1,2,\ldots,n$.
\STATE 2. Fit the constrained model corresponding to $H_{0}$ (the null) and estimate $\alpha,\beta(t)$ from the constrained minimization criteria in (\ref{sofroptorig}), denote the estimates $\hat{\alpha}_c, \hat{\beta}_{c}(t)$.
\STATE 3. Compute test statistic $T$ (\ref{boot}) based on the null and full model fits, denote this as $T_{obs}$.
\STATE 4. Resample B sets of bootstrap residuals $\{e^*_{b,i}\}_{i=1}^{n}$ from residuals $\{e_{i}\}_{i=1}^{n}$ obtained in step 1.
\STATE 5. for $b = 1$ to $B$ 
\STATE 6. Generate scalar response under the constrained null model as
$$Y^*_{b,i}=\hat{\alpha}_c+ \int_{T} X_i(t)\hat{\beta}_c(t)dt+e^*_{b,i}.$$
\STATE 7. Given the bootstrap data set $\{X_i(t),Y^*_{b,i}\}_{i=1}^{n}$ fit the null and the full model to compute the test statistic $T^*_b$.
\STATE 8. end for
\STATE 9. Calculate the p-value of the test as $\hat{p}=\frac{\sum_{b=1}^{B} I(T^*_b \geq T_{obs})}{B}$.
\end{algorithmic}
\end{algorithm}

\subsubsection*{Shape Testing with Functional Response}
We now consider the models with functional response such as the FOSR model (\ref{FOSR1}) or the FLCM (\ref{FLCM1}). Here again, we want to test, $$H_{0} : \beta_1(\cdot) \in \mathcal{F}\cap\mathcal{H} \hbox{ \;\; versus \;\; } H_{1} :\beta(\cdot)\in \mathcal{H},$$ where $\mathcal{F}$ is the specific class of shape restricted functions. The testing method is based on a similar bootstrap procedure as in the case with scalar response. While performing testing, we don't enforce pre-whitening in the step 2 of our estimation, as the estimated residuals from the unconstrained model 
(\ref{func response}) (corresponding to the estimators obtained via minimizing $\sum_{i=1}^{n}||{\*Y_i}-\^B_0\bm\beta_0-\^W_i\bm\beta_1||_2^{2}$) asymptotically has the same covariance as the original residuals and bootstrapping these generates residuals from the same covariance structure without going into the need for estimating it. The detailed testing procedure is presented in Algorithm 1 within Appendix C of the Supplementary Material.

\subsection{Selection of order of the Bernstein Polynomial Basis}
The order of the Bernstein polynomial basis $N$ controls the smoothness of the regression coefficient functions $\beta(t)$. A smaller $N$ might introduce bias in estimation, while a larger $N$ can make the coefficient functions wiggly.
We follow a truncated basis approach  \citep{Ramsay05functionaldata,fan2015functional}, by restricting the number of BP basis functions to ensure the estimated regression coefficient function is smooth. The empirically optimal number of basis functions is chosen in a data-driven way \citep{ahkim2017shape} via $V$-fold ($V=5$ in this article) cross-validation method \citep{wang2012shape} using cross-validated residual sum of squares for both the scalar and functional response regression. In particular, the cross-validated residual sum of square for the scalar response case is defined as follows
 $$CV_{s}(N)= \sum_{v=1}^{V}\sum_{i=1}^{n_v}(Y_{i,v}-\hat{Y}_{i,v,N}^{-v})^{2}. $$ 
 Here $\hat{Y}_{i,v,N}^{-v}$ is the fitted value of the scalar outcome $Y_{i,v}$, within the $v^{th}$ fold obtained by applying a model trained on the rest $(V-1)$ folds using Bernstein polynomials of order $N$. Similarly for the functional response case, cross-validated residual sum of square is defined as follows
  $$CV_{f}(N)= \sum_{v=1}^{V}\sum_{i=1}^{n_v}||\*Y_{i,v}-\hat{\*Y}_{i,v,N}^{-v}||_2^{2}. $$
 The empirically optimal $N$ is then chosen based on a grid search as the minimizer of $CV_{s}(N)$ or $CV_{f}(N)$. It should however be noted that there is perhaps no universal (data-dependent) method of empirically selecting the tuning parameter $N$, even when one can establish a sharp rate of asymptotic convergence.  
\section{Simulation Studies}
\label{simul}
In this Section, we investigate the performance of the proposed estimation method under shape constraints via simulations. To this end, the following scenarios are considered.
\subsection{Data Generating Scenarios}
\subsection*{Scenario A: SOFR, non-negative constraint}
We generate data from the scalar-on-function regression (SOFR) model given by \begin{equation*}
    Y_i=\alpha+ \int_{T} X_i(t)\beta(t)dt+\epsilon_i,
\end{equation*}
where $\alpha=0.15$ and $\beta(t)=0.1*sin (\pi t)$. The residuals $\epsilon_i\sim N(0,0.05^2)$ (i.i.d). We consider a dense design with $m=50$ equispaced time-points in $\mathcal{T}=[0,1]$ and sample size $n\in \{25,50,100\}$. The covariate process $X_i(t)$ is generated as $X_i(t)=\sum_{k=1}^{20}\psi_{ik}\phi_k(t)$, where $\phi_k(t)$ are orthogonal basis polynomials (of degree $k-1$) and $\psi_{ik}$ are  mean zero and independent Normally distributed scores with variance $\sigma^2_k=(20-k+1)$. We consider estimation in the above model under the constraint $\beta(t)\geq0$.

\subsection*{Scenario B: FLCM, non-increasing constraint}
We generate data from the functional linear concurrent model (FLCM)
\begin{equation*}
    Y_i(t)=\beta_0(t)+X_i(t)\beta_1(t) +\epsilon_i(t),
\end{equation*}
where the coefficient functions are given by $\beta_0(t)=8sin(\pi t)$, $\beta_1(t)=5cos(\pi t)$.  The covariate process $X_i(t)$ is generated as $X_i(t)=\sum_{k=1}^{5}\psi_{ik}\phi_k(t)$, where $\phi_k(t)$ are orthogonal basis polynomials (of degree $k-1$) and $\psi_{ik}$ are  mean zero and independent Normally distributed scores with variance $\sigma^2_k=(5-k+1)$. The error process $\epsilon_{i}(t)$ is generated as 
$\epsilon_i(t)=\xi_{i1} cos(t) +\xi_{i2}sin(t) + N(0,0.5^2),$
where $\xi_{i1}\stackrel{iid}{\sim}\mathcal{N}(0, 0.5^2)$  and $\xi_{i2}\stackrel{iid}{\sim}\mathcal{N}(0,0.75^2)$. We consider a dense design with $m=40$ equispaced time-points in $\mathcal{T}=[0,1]$ and sample size $n\in\{25,50,100\}$. We consider estimation in the above model under the constraint $\beta_1(t)$ is decreasing.

\hspace{- 8 mm}
\textbf{Additional Simulations:}
We consider a sparse design under the above data generating set up, where the functional covariate $X_i(t)$ and the functional response $Y_i(t)$ are observed over randomly chosen $m_i$ time-points ($m_i\sim Unif\{5,6,\ldots,10\}$) from the dense grid of $40$ equispaced time-points in $\mathcal{T}=[0,1]$. Sample size $n\in\{50,100\}$ is considered for this sparse scenario.

\subsection*{Scenario C: FLCM, non-decreasing and concave constraint}
We generate data from another FLCM given by,
\begin{equation*}
    Y_i(t)=\beta_0(t)+X_i(t)\beta_1(t) +\epsilon_i(t),
\end{equation*}
where the coefficient functions are given by $\beta_0(t)=3cos(\pi t)$, $\beta_1(t)=5sin(\frac{\pi}{2} t)$.  The covariate process $X_i(t)$ and the error process $\epsilon_{i}(t)$ is generated exactly as in scenario B. We  again consider a dense design with $m=40$ equispaced time-points in $\mathcal{T}=[0,1]$ and sample size $n\in\{25,50,100\}$. We consider estimation in the above model under the constraint $\beta_1(t)$ is non-decreasing, and $\beta_1(t)$ is concave.

We consider 200 Monte-Carlo (M.C) replications from the above specified  simulation scenarios to assess the performance of the proposed estimation method.


\subsection{Summary of Results from Simulated Data Scenarios}
\subsection*{Performance under scenario A:}
\begin{figure}[ht]
\centering
\includegraphics[width=.9\linewidth , height=.5\linewidth]{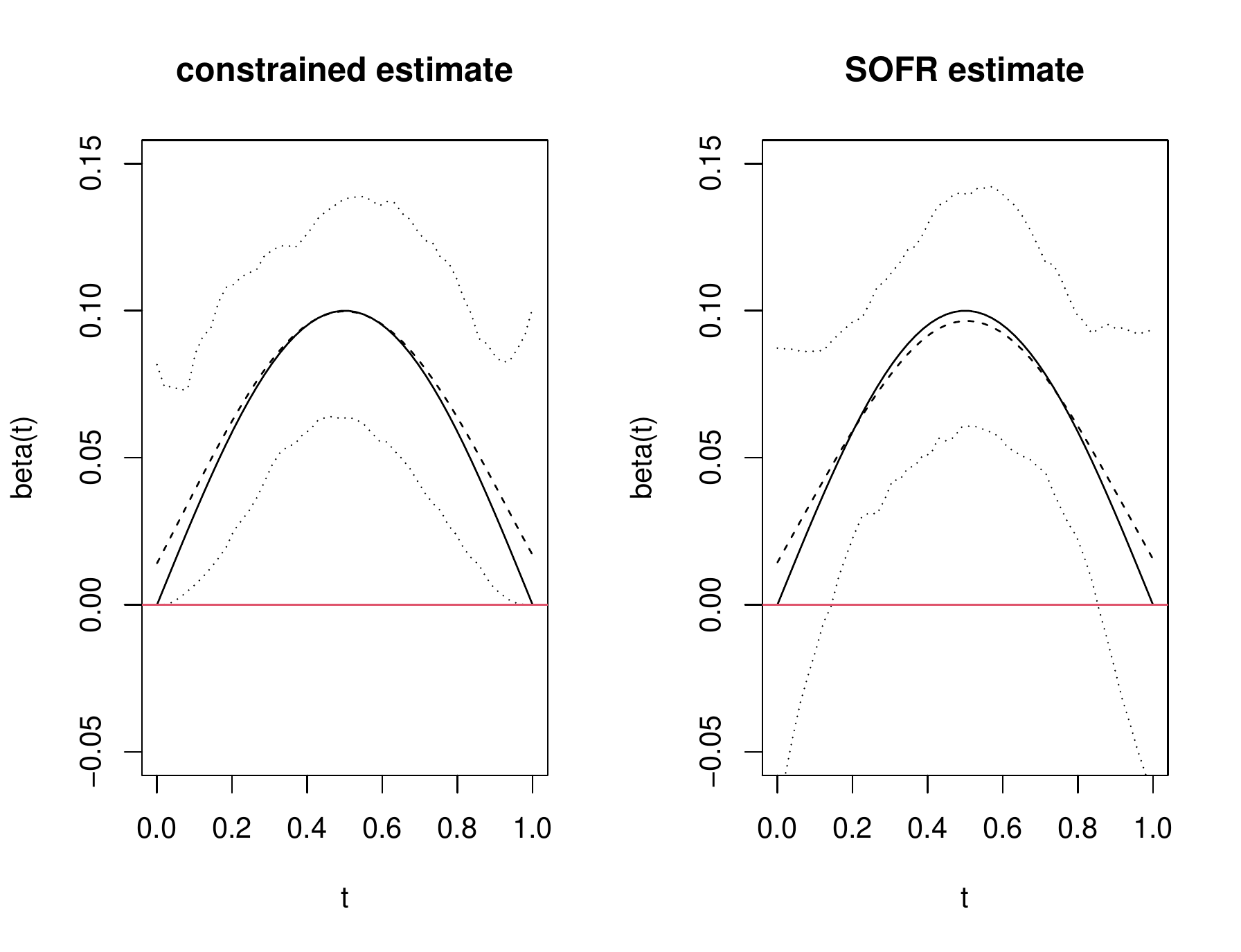}
\caption{Estimated coefficient function (dashed line) and true coefficient function (solid line) along with $95\%$ point-wise confidence interval (dotted lines) from the constrained (left) and unconstrained (right) method, simulation scenario A, n=50. }
\label{fig:fig1}
\end{figure}

We consider estimation in the SOFR model of scenario A under the constraint $\beta(t)\geq0$ (satisfied by the true coefficient function). The performance of the proposed constrained method is compared with existing unconstrained approach using the standard \texttt{"pfr"} function for SOFR within the \texttt{refund} package in R. Figure \ref{fig:fig1} displays the estimated coefficient function $\hat{\beta}(t)$ (for the sample size $n=50$) averaged over the 200 M.C replications from both the constrained and the unconstrained method. On average, the order of Bernstein-polynomials chosen by five-fold  ($V=5$) cross-validation was $N=4$ across the three sample sizes. The $95\%$ point-wise confidence intervals of the coefficient function are also shown based on the M.C replications. We can notice the confidence bands for the constrained method does not include zero for any $t$, and also produce closer estimates to the true function compared to the unconstrained approach. The confidence intervals are also narrower, specially at the boundaries compared to the ones from the unconstrained method. The average M.C mean square error (IMSE) of the estimated coefficient function $\hat{\beta}(t)$ defined as IMSE $= \int_{\mathcal{T}}(\hat{\beta}(t)-\beta(t))^2dt $, from both the constrained and unconstrained method, are reported in Table \ref{tab:my-table1}.
We can notice that the constrained estimates produce smaller average IMSE compared to the unconstrained estimates. In particular, based on Table \ref{tab:my-table1}, the constrained estimators, on an average, are found to be $33\%$ more efficient compared to the unconstrained estimates in terms of average IMSE. As the sample size increase, the IMSE from both the methods become negligible indicating consistency of the estimators.

\begin{table}[ht]
\centering
\begin{tabular}{|c|c|c|c|}
\hline
Sample size (n) & Constrained method & Unconstrained method & P-value              \\ \hline
25              & 0.9 (1.0)    & 1.3 (1.1)      & 2.99$\times 10^{-5}$               \\ \hline
50              & 0.4 (0.3)    & 0.6 (0.4)      & 1.5$\times 10^{-6}$ \\ \hline
100             & 0.2 (0.2)    & 0.3 (0.2)      & 0.0004               \\ \hline
\end{tabular}
\caption{Average integrated mean square error  ($\times$ 1000) over 200 Monte-Carlo replications, scenario A. Standard errors of IMSE   ($\times$ 1000) are reported in the parenthesis. P-values are obtained from two sample t-test.}
\label{tab:my-table1}
\end{table}

\subsubsection*{Projection-based confidence intervals}
We apply the projection based method outlined in Algorithm \ref{algo 12} to obtain point-wise asymptotic $95\%$ confidence interval of the regression coefficient function $\beta(t)$ under the shape constraint $\beta(t)>0$. Table S1 in Supplementary Material reports the average empirical coverage of the confidence intervals across the three sample sizes for a range of choices for the order of the Bernstein polynomial basis, $N$. Average estimated coverages are close to the nominal $95\%$ coverage for $N\geq4$, which is the average order of the Bernstein polynomial basis chosen by our proposed cross-validation criterion. The average width of the confidence interval is  found to be comparable or smaller than the unconstrained (\texttt{"pfr"}) method for choices of $N$ around $N_0=4$. The projection based confidence interval for a particular replication ($n=100$) is displayed in Figure S1 in Supplementary Material.

\subsubsection*{Testing shape constraints}
We consider testing the following shape constraints
for the simulation scenario A. i) $H_{0} : \beta_1(t) \geq 0  \hbox{ for all $t\in [0,1]$}$, ii) $H_{0} : \beta_1(t) \hspace{2 mm} \textit{is concave} \hbox{ for all $t\in [0,1]$}$ iii) $H_{0} : \beta_1(t) \hspace{2 mm} \textit{is increasing}$ $\hbox{for all $t\in [0,1]$}$. The true coefficient function in this scenario is $\beta(t)=0.1*sin (\pi t)$ which satisfy i) and ii) but does not satisfy iii). Table \ref{tab:my-table1_test} displays the rejection rates of the bootstrap test for the three null hypothesis and three sets of sample sizes for the nominal level of $\alpha=0.05$.

\begin{table}[ht]
\centering
\begin{tabular}{cccc}
\hline
$H_{0}$ & $n=25$ & $n=50$  & $n=100$             \\ \hline
$ \beta_1(t) \geq 0$             & 0.055    & 0.05     & 0.065               \\ \hline
$\beta_1(t) \hspace{2 mm} \textit{is concave}$            & 0.04     & 0.07      & 0.055 \\ \hline
$\beta_1(t) \hspace{2 mm} \textit{is increasing}$             & 0.235     & 0.55       & 0.84               \\ \hline
\end{tabular}
\caption{Rejection rates for the respective hypothesis from the bootstrap based test based on 200 M.C simulations from Scenario A.}
\label{tab:my-table1_test}
\end{table}
We notice the rejection rates remain close to the nominal level of $0.05$ when the null hypothesis is true (case i and ii) and is increasing to 1 as the sample size increase when the null is false (case iii) indicating consistency of the proposed testing method.

\subsection*{Performance under scenario B:}
\begin{figure}[ht]
\centering
\includegraphics[width=1\linewidth , height=.45\linewidth]{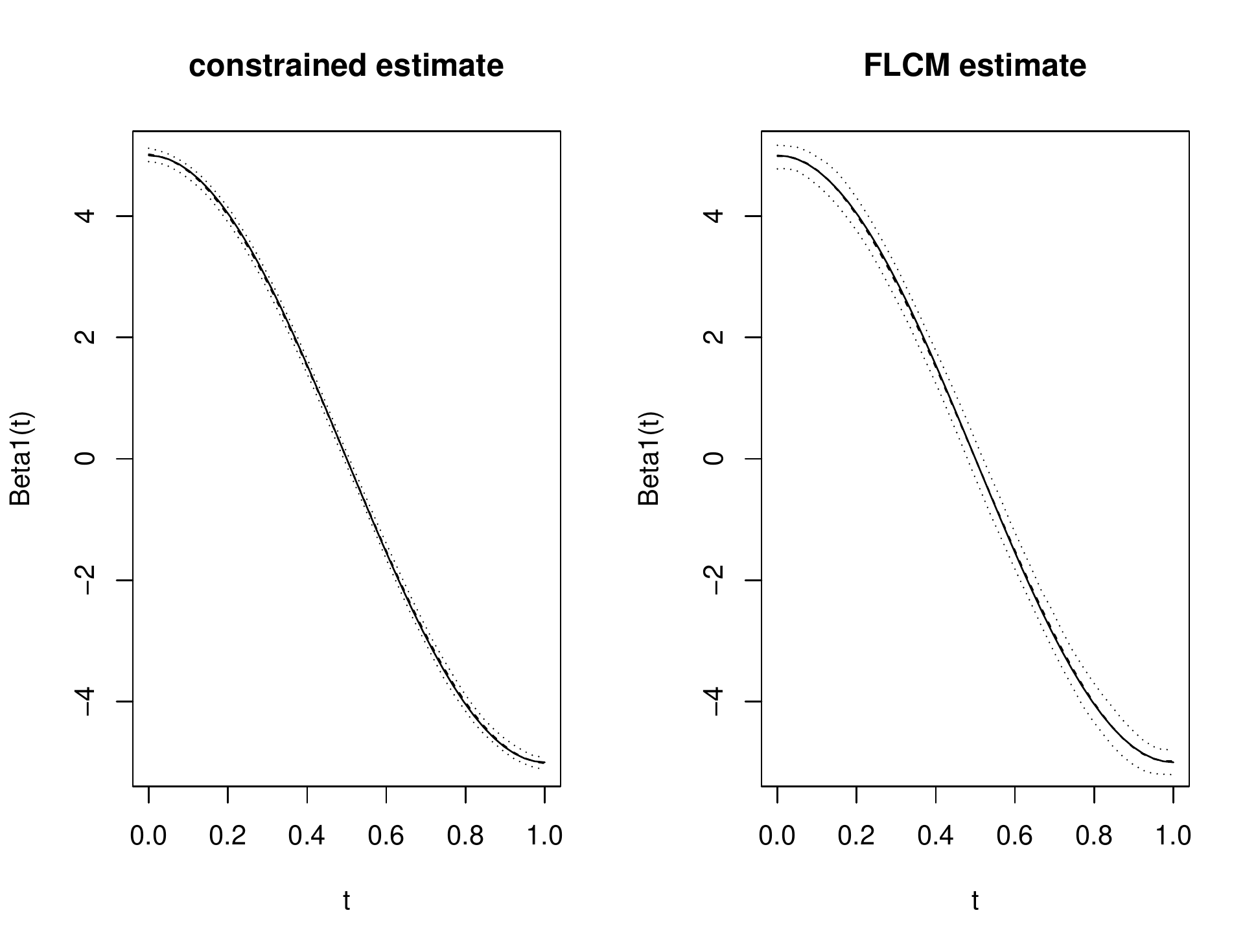}
\caption{Estimated coefficient function (dashed line) and true coefficient function (solid line) along with $95\%$ point-wise confidence interval (dotted lines) from the constrained (left) and unconstrained (right) method, simulation scenario B, n=50. }
\label{fig:fig2}
\end{figure}

We consider estimation in the functional linear concurrent model of scenario B under the constraint $\beta_1(t)$ is decreasing. The average order of Bernstein-polynomial basis chosen by five-fold ($V=5$) cross-validation was $N=5$, across the three sample sizes. Figure \ref{fig:fig2} displays the estimated coefficient function $\beta_1(t)$ (for sample size $n=50$) averaged over 200 M.C replications from both the constrained and unconstrained approach (smoothing spline implemented using \texttt{"pffr"} function within the \texttt{refund} package) along with their $95\%$ point-wise confidence intervals based on the Monte-Carlo replications. 

It can be noticed the constrained approach produce much narrower confidence interval of the estimate indicating lower uncertainty of the estimated function in the restricted parameter space.
The M.C average mean square error (IMSE) (multiplied by 100) of the estimated coefficient function $\hat{\beta}_1(t)$ from the constrained and unconstrained method is reported in Table \ref{tab:my-table2r}. It can be observed that the proposed constrained estimates have much smaller average IMSE compared to the unconstrained estimates. Across the sample sizes, the constrained estimates are found to be $346\%$ more efficient compared to the unconstrained estimators. As the sample size increase, the IMSE from both the methods again become negligible indicating consistency of the estimators. The IMSE (multiplied by 100) of the estimated coefficient function $\hat{\beta}_1(t)$ from the sparse design setup is reported in Table S3 of the Supplementary Material, where similar improvement in efficiency can be noticed from the constrained approach.

\begin{table}[ht]
\centering
\begin{tabular}{|c|c|c|c|}
\hline
Sample size (n) & Constrained method & Unconstrained method & P-value              \\ \hline
25              & 1.23 (1.14)    & 5.37 (5.15)      & $<2.2\times 10^{-16}$               \\ \hline
50              & 0.46 (0.30)    & 2.13 (1.56)      & $<2.2\times 10^{-16}$ \\ \hline
100             & 0.26 (0.15)    & 1.14 (0.85)      & $<2.2\times 10^{-16}$               \\ \hline
\end{tabular}
\caption{Average integrated mean square error ($\times$ 100) over 200 Monte-Carlo replications, scenario B. Standard errors of IMSE  ($\times$ 100) are reported in the parenthesis. P-values are obtained from two sample t-test.}
\label{tab:my-table2r}
\end{table}

\subsubsection*{Projection-based confidence intervals}
We display the projection-based point-wise asymptotic $95\%$ confidence interval of the regression coefficient function $\beta_1(t)$ under the shape constraint: $\beta_1(t)$ is decreasing. Table S2 in Supplementary Material reports the average empirical coverage of the confidence intervals across the three sample sizes for a range of choices of $N$. The estimated coverage is close to the nominal $95\%$ coverage for $N\geq 5$, which is the average order of the Bernstein polynomial basis chosen by our proposed cross-validation criterion. The average width of the confidence interval is found to be smaller than the unconstrained (\texttt{"pffr"}) method for choices of $N$ around $N_0=5$, while yielding the correct coverage. The projection-based confidence interval of $\beta_1(t)$ for a particular replication ($n=100$) is displayed in the Supplementary Figure S2.

\subsubsection*{Testing shape constraints}
In this scenario, We test the following shape constraints. i) $H_{0} : \beta_1(t) \hspace{2 mm} \textit{is decreasing}$ $\hbox{$\forall$ $t\in [0,1]$}$, ii) $H_{0} : \beta_1(t) \hspace{2 mm} \textit{is convex} \hbox{ $\forall$ $t\in [0,1]$}$ iii) $H_{0} : \beta_1(t) \hspace{2 mm} \textit{is concave}$ $\hbox{$\forall$ $t\in [0,1]$}$. The true coefficient function in this scenario is $\beta_1(t)=5cos(\pi t)$ which satisfies i) but does not satisfy ii) and ii). Table \ref{tab:my-table2test} displays the rejection rates of the bootstrap test for the three null hypothesis and three sets of sample sizes for the nominal level of $\alpha=0.05$.
\begin{table}[H]
\centering
\begin{tabular}{cccc}
\hline
$H_{0}$ & $n=25$ & $n=50$  & $n=100$             \\ \hline
$\beta_1(t) \hspace{2 mm} \textit{is decreasing}$           & 0.06    & 0.07     & 0.065               \\ \hline
$\beta_1(t) \hspace{2 mm} \textit{is convex}$            & 1     & 1      & 1 \\ \hline
$\beta_1(t) \hspace{2 mm} \textit{is concave}$             & 1     & 1       & 1               \\ \hline
\end{tabular}
\caption{Rejection rates for the respective hypothesis from the bootstrap based test based on 200 M.C simulations from Scenario B.}
\label{tab:my-table2test}
\end{table}
We again observe that the rejection rates remain close to the nominal level of $0.05$ when the null hypothesis is true (case i) and is 1 for all the sample sizes when the null is false (case ii and iii) indicating satisfactory power of the proposed testing method.

\hspace*{-8 mm}
\textit{Remark 4:}\\
Supplementary Simulation Scenario S1 illustrates that the proposed estimation approach and the projection based asymptotic confidence interval is able to capture the true coefficient function accurately, even when it lies on the boundary of the parameter space (e.g., constant function under decreasing constraint).

\subsection*{Performance under scenario C:}
\begin{figure}[ht]
\centering
\includegraphics[width=1\linewidth , height=.6\linewidth]{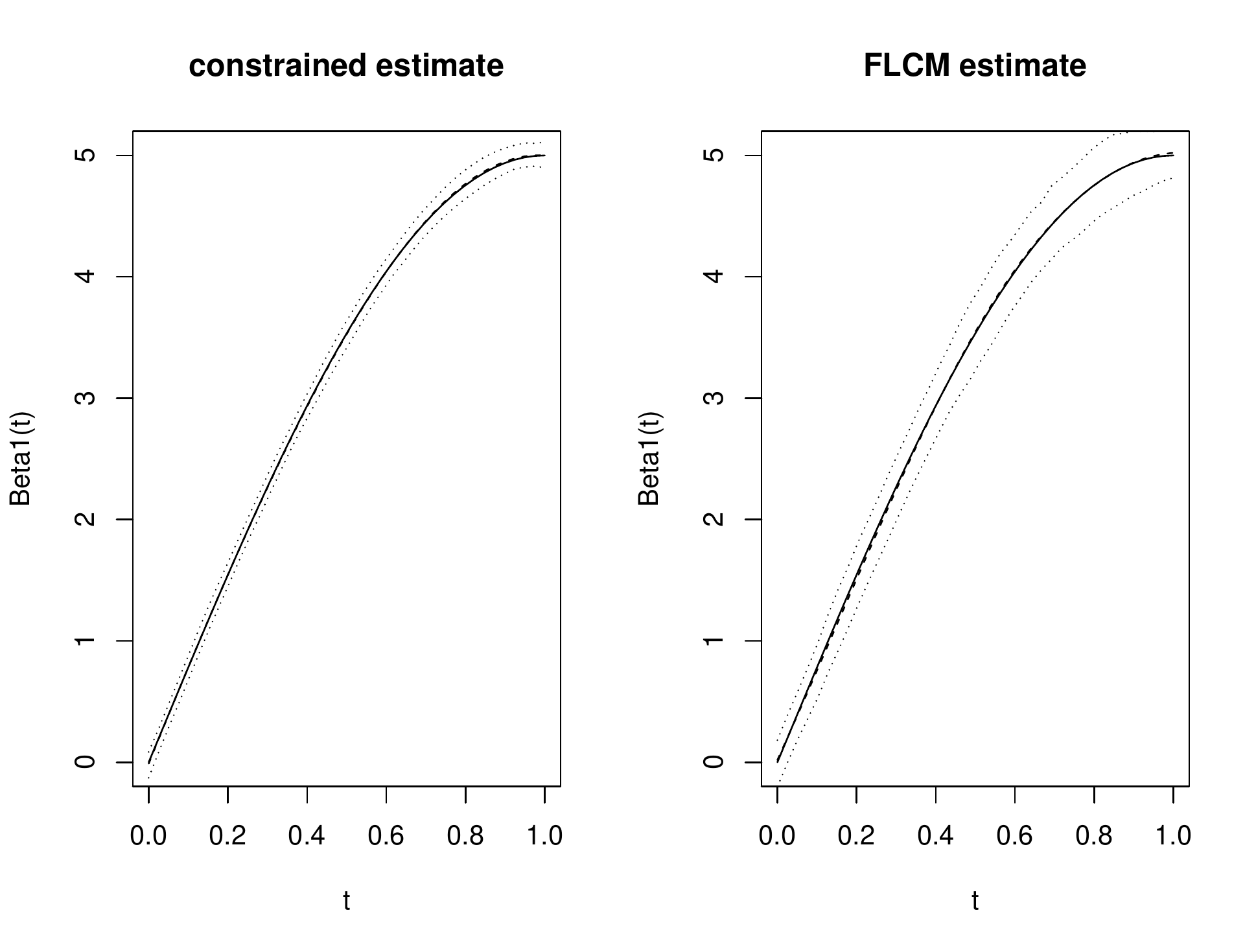}
\caption{Estimated coefficient function (dashed line) and true coefficient function (solid line) along with $95\%$ confidence interval (dotted lines) from the constrained (left) and unconstrained (right) method, simulation scenario C, n=50. }
\label{fig:fig3}
\end{figure}

We consider estimation in the FLCM described in simulation scenario C under the constraints $\beta_1(t)$ is increasing and $\beta_1(t)$ is concave. The average order of Bernstein-polynomial basis chosen by five-fold ($V=5$) cross-validation was $N=5$ across the three sample sizes.
Figure \ref{fig:fig3} displays the estimated coefficient function $\beta_1(t)$ averaged over 200 M.C replications from both the constrained and unconstrained approach along with their $95\%$ point-wise confidence intervals.
It can again be noticed the constrained approach produce much narrower confidence interval of the estimated coefficient function $\hat{\beta}_1(t)$ indicating lower uncertainty of the estimate in the restricted parameter space. The average M.C mean square error (IMSE) of the estimated coefficient function $\hat{\beta}_1(t)$ from both the constrained and unconstrained method is reported in Table \ref{tab:my-table3r}.
\begin{table}[ht]
\centering
\begin{tabular}{|c|c|c|c|}
\hline
Sample size (n) & Constrained method & Unconstrained method & P-value              \\ \hline
25              & 9.5 (9.8)    & 49.3 (49.7)      & $<2.2\times 10^{-16}$               \\ \hline
50              & 3.1 (2.6)    & 19.5 (15.1)      & $<2.2\times 10^{-16}$ \\ \hline
100             & 1.4 (1.1)    & 10.5 (8.2)      & $<2.2\times 10^{-16}$               \\ \hline
\end{tabular}
\caption{Average integrated mean square error ($\times$ 1000) over 200 Monte-Carlo replications, scenario C. Standard errors of IMSE ($\times$ 1000) are reported in the parenthesis. P-values are obtained from two sample t-test.}
\label{tab:my-table3r}
\end{table}
We again observe that the proposed constrained estimates have much smaller average IMSE compared to the unconstrained estimates. Specifically, the constrained estimates are found to be $485\%$ more efficient compared to the unconstrained estimator. As the sample size increase, the IMSE from both the methods again become asymptotically negligible.

The simulation results in this section illustrate the advantages of the proposed estimation method in functional regression models under shape restrictions. For smaller sample sizes, such shape restrictions lead to reduced uncertainty of the coefficient functions in the restricted parameter space.

\section{Real Data Applications}
\label{realdata}
In this section, we demonstrate applications of the proposed shape constraint estimation method in functional regression models. First, we consider a mental health schizophrenia collaborative
study analyzing evolution of drug effect on the severity of illness.
Next, we apply the proposed estimation method for distributional analysis of quantile-functions of physical activity data from  
Baltimore Longitudinal Study of Aging (BLSA).
\subsection{Application 1: Mental Health Schizophrenia Collaborative
Study}
We consider data from the National Institute of Mental Health Schizophrenia Collaborative Study used in \cite{ahkim2017shape}. The response of interest is severity of illness measured in the Inpatient Multidimensional Psychiatric Scale (IMPS), ranging from 1 (normal) to 7 (among the most extremely ill). The patients ($n=437$) considered in this study were randomly assigned to a treatment (drug) or placebo and measured at weeks $0,1,2,\ldots,6$. Majority of the patients were measure at weeks $0,1,3,6$ with a few being additionally measured on weeks $2,4,5$. The primary interest here is to assess the efficacy of the drug. We consider a function-on-scalar regression model,
\begin{equation}
    Y_i(t)=\beta_0(t)+G_i\beta_1(t)+\epsilon_i(t), \label{schrizo}
\end{equation}
where $Y_i(t)$ denotes disease severity at week $t$ for subject $i$ and $G_i$ is a indicator of the treatment group for the subject ($G_i=1$, if subject $i$ received drug). 
\begin{figure}[ht]
\begin{center}
\includegraphics[width=.7\linewidth , height=.7\linewidth]{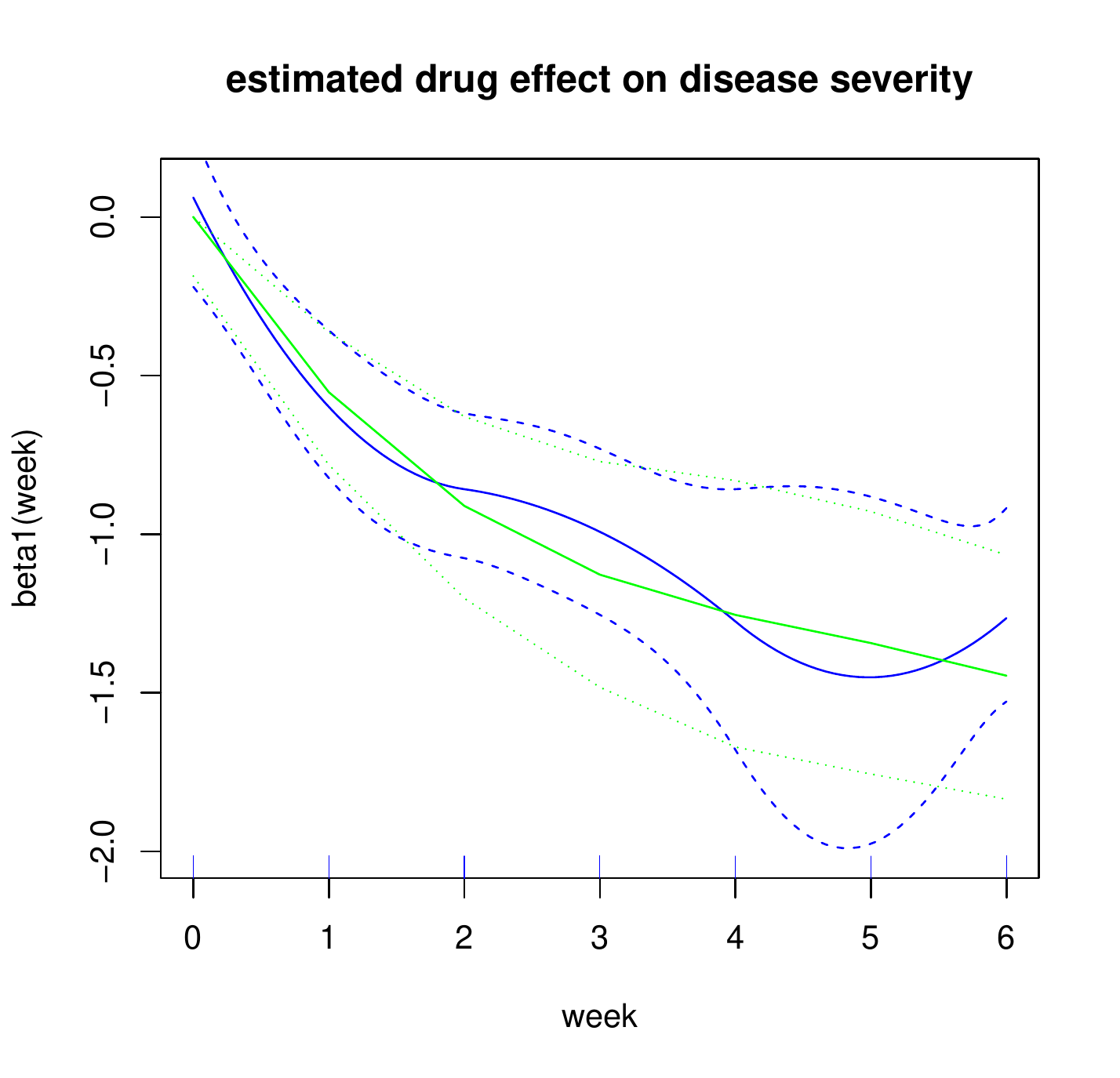} 
\end{center}
\caption{Estimated effect of drug on disease severity as function of week. Constrained ($\beta_1(t)\leq0$) estimate (green solid line) and unconstrained estimate (blue solid line), along with their $95\%$ confidence intervals (dotted for constrained and dashed for unconstrained).}
\label{fig:figsc1}
\end{figure}
The time-varying coefficient function $\beta_1(t)$ captures the effect of the drug on disease severity. A negative $\beta_1(t)$ would prove the effectiveness of the drug while a negative and decreasing $\beta_1(t)$ would suggest the magnitude of the effectiveness of the drug increase as the weeks progress.

We apply the proposed residual bootstrap-based test in Section 2, and test the constraints i) $\beta_1(t)\leq0$ and ii) $\beta_1^{'}(t)\leq0$, i.e., $\beta_1(t)$ is decreasing. The P-values of the tests are calculated to be 0.53, 0.6 respectively. Hence we fail to reject the hypothesis that the effect of the drug is negative and the effect is decreasing, which matches with the findings by \cite{ahkim2017shape}. Next, we apply the proposed estimation method in this article under the shape constraint $\beta_1(t)\leq0$, with the prior knowledge that the drug is effective \citep{ahkim2017shape} (ascertained by our proposed test). The degree of the Bernstein polynomial used to model the coefficient functions was chosen to be $3$ by five-fold cross-validation indicating sufficiency of a cubic fit. The estimated coefficient function $\beta_1(t)$ capturing the effect of the drug on disease severity is shown in Figure \ref{fig:figsc1}. The estimated coefficient function from an unconstrained fit using penalized function-on-scalar regression (obtained using \texttt{"pffr"} function within the \texttt{refund} package in R) is also displayed. The estimates are accompanied with their respective $95\%$ confidence intervals. The average width of the confidence intervals from the constrained method ($0.62$) is found to be smaller compared to that of the unconstrained method ($0.65$). 

We notice a negative and mostly decreasing $\beta_1(t)$, illustrating the effectiveness of the drug which is captured by both the constrained and the unconstrained estimator.
\begin{figure}[ht]
\begin{center}
\includegraphics[width=.6\linewidth , height=.6\linewidth]{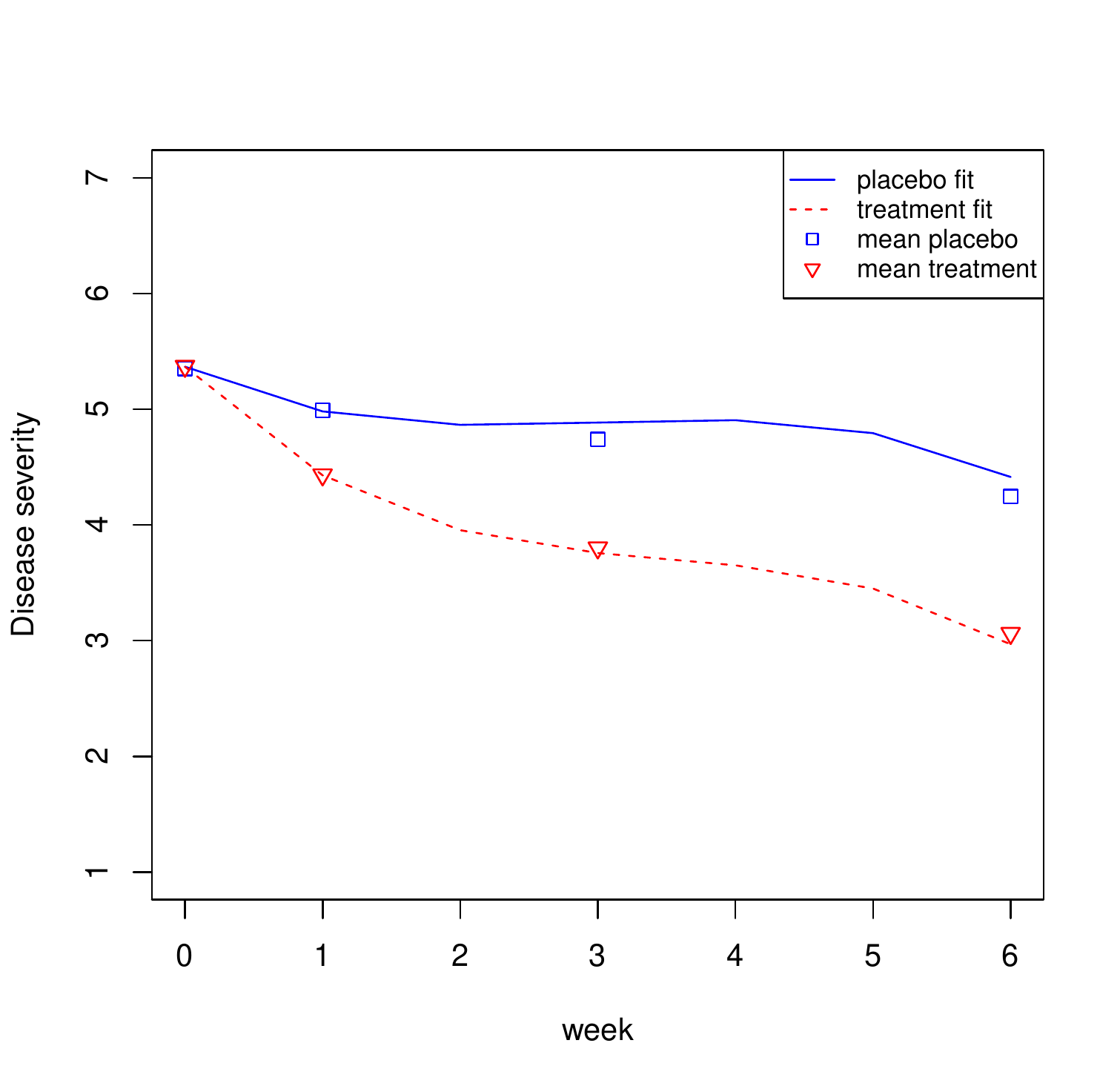} 
\end{center}
\caption{Fitted disease severity trajectories for placebo (solid line) and treatment group (solid line), Schizophrenia data.}
\label{fig:figsc2}
\end{figure}
The fitted disease severity trajectories from the constrained method is shown in Figure \ref{fig:figsc2} for the treatment and the placebo group, which are very close to the average observed values of disease severity (IMPS).

\subsection{Quantile Function on Scalar Regression (QFOSR) of Physical Activity Data from BLSA}
As our second example, we model subject-specific distributions of accelerometry-measured physical activity from Baltimore Longitudinal Study of Aging (BLSA), the longest-running scientific study of aging in the United States. Specifically, we are interested in how the subject-specific quantile functions of minute-level activity counts are associated with age, gender, height and weight. We use a quantile function-on-scalar regression framework introduced in \cite{yang2020quantile} for distributional analysis of physical activity. Prior studies on BLSA \citep{xiao2015quantifying} have focused on modeling diurnal variability of physical activity. Activity counts were measured using a chest-worn Actiheart physical activity monitor on participants in their free-living environment for several consecutive days. For this analysis, we consider a sample of $n=857$ subjects in BLSA and a single visit of each participant. Table \ref{tab:my-table4} presents the descriptive
statistics of the sample. 

\begin{figure}[ht]
\begin{center}
\includegraphics[width=.6\linewidth , height=.6\linewidth]{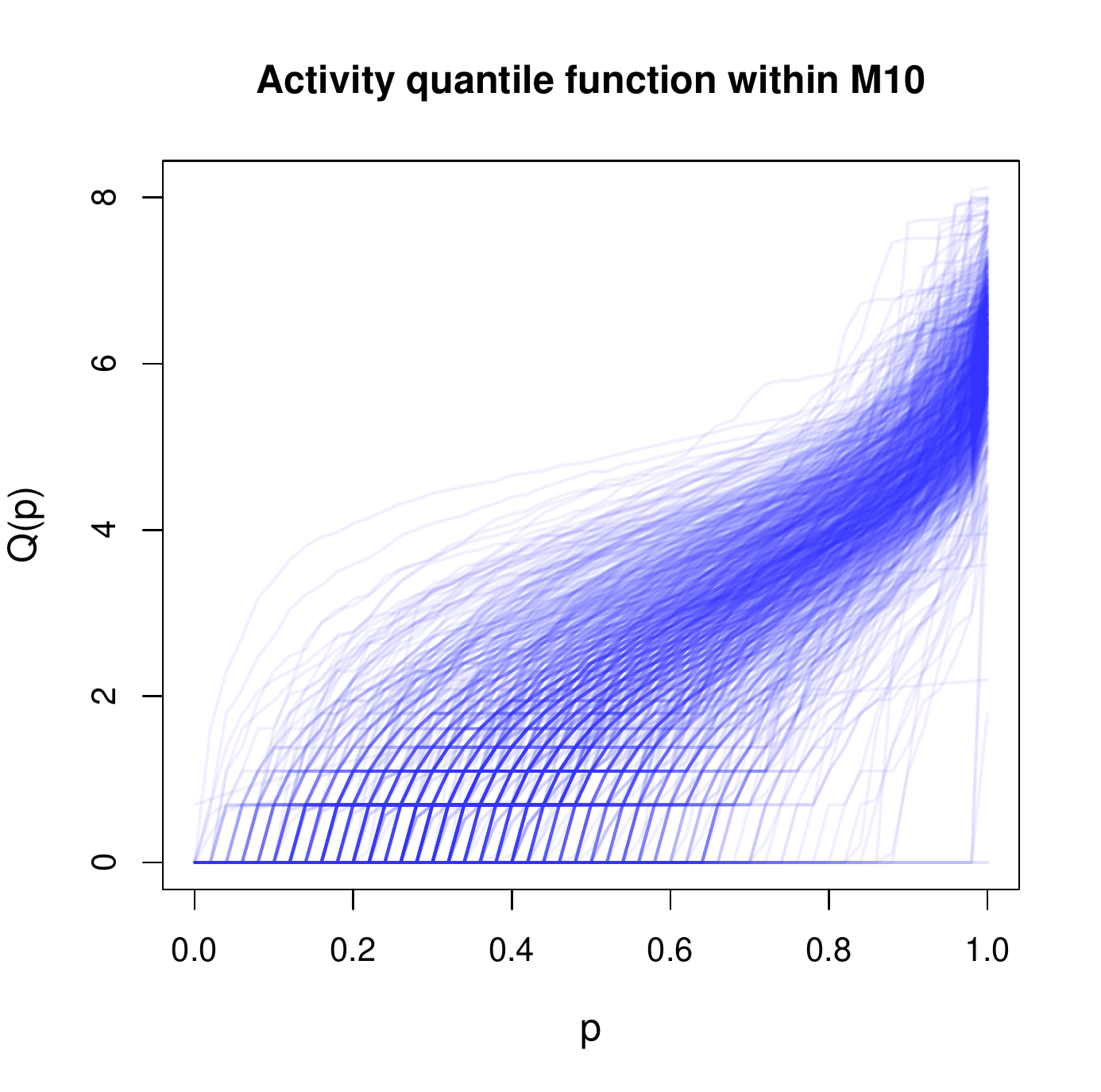} 
\end{center}
\caption{Subject-specific quantile functions of log-transformed physical activity in their most active 10 hour window $M_{10}^{i}$.}
\label{fig:figbl1}
\end{figure}

\begin{table}[H]
\centering
\small
\begin{tabular}{cccccccc}
\hline
Characteristic     & \multicolumn{2}{c|}{Complete (n=857)} & \multicolumn{2}{c|}{Male (n=420} & \multicolumn{2}{c|}{Female (n=437)} & P value         \\ \hline
                   & Mean          & SD  & Mean     & SD      & Mean/Freq       & SD        &                 \\ \hline
Age                & 66.83               & 13.17           & 68.11         & 13.36    & 65.6           & 12.88      & 0.005            \\ \hline
Height (m) & 1.69               & 0.09           & 1.76         & 0.07    & 1.63           & 0.06      & $<2.2 \times 10^{-16}$          \\ \hline
Weight (Kg)                & 78.33               & 16.37           & 85.02         & 14.92    & 71.90            & 15.09      & $<2.2 \times 10^{-16}$          \\ \hline
\end{tabular}
\caption{Summary statistics for the complete, male and female samples considered for the BLSA analysis.}
\label{tab:my-table4}
\end{table}
Subject-specific daily PA is represented via 1440 minute-level activity counts. For the analysis in this paper, we limit our attention to data collected on Mondays for each subject visit. We also only consider activity counts from participants in their most active 10 hour period ($M_{10}^{i}$) \citep{witting1990alterations}, since this can serve as a proxy period for most of daily physical activity. The activity counts are log transformed using the transformation $A\longrightarrow log(A+1)$ to remove possible skewness in the data.
Subsequently, we encode subject level physical activity data in the window $M_{10}^{i}$ (depends on the subject $i$) using subject-specific quantile functions $Q_i(p)$ for $i=1,2,
\ldots,n=857$, and $p\in [0,1]$. Figure \ref{fig:figbl1} displays the observed (empirical) subject-specific quantile functions $Q_i(p)$.

We model the subject-day specific quantile functions as outcomes using the quantile function-on-scalar regression model as follows:
\begin{equation}
    Q_i(p)=\beta_0(p)+x_{age,i}\beta_{age}(p)+x_{sex,i}\beta_{sex}(p)+x_{height,i}\beta_{H}(p)+x_{weight,i}\beta_{W}(p) +\epsilon_i(p). \label{qfosr}
\end{equation}
The functional regression coefficients $\beta_{age}(p)$, $\beta_{sex}(p)$, $\beta_{H}(p)$ and $\beta_{W}(p)$ capture the effects of age, sex (Male = 1, Female = 0), height, and weight on the quantile level $p$ of subject-specific physical activity. The intercept function $\beta_0(p)$ represents the baseline $p-th$ quantile level of physical activity. \cite{yang2020quantile} proposed to use  quantlets, data-driven basis functions, for estimating the functional regression coefficients in QFOSR. 
The quantlet based estimation approach does not explicitly impose monotonicity in the predicted quantile functions, although the observed (empirical) quantile functions are monotone \citep{yang2020quantile}, and most of the predicted quantile functions in their applications were found to be monotone and non-decreasing.

\cite{yang2020random} proposed a non-decreasing basis based estimation using I-splines \citep{ramsay1988monotone}
or Beta CDFs which enforce this monotonicity at the estimation step. This produces the coefficient functions of form $\beta_{a}(p)=\sum_{k=1}^{K}\Psi_k(p)\hat{\beta}_{ak}$, where $\hat{\beta}_{ak} \geq0$. Notice that this essentially enforces the coefficient functions of the scalar predictors to be monotone and non-decreasing which might not be necessarily true for many of the covariates. 

\begin{figure}[H]
\begin{center}
\begin{tabular}{ll}
\includegraphics[width=.38\linewidth , height=.38\linewidth]{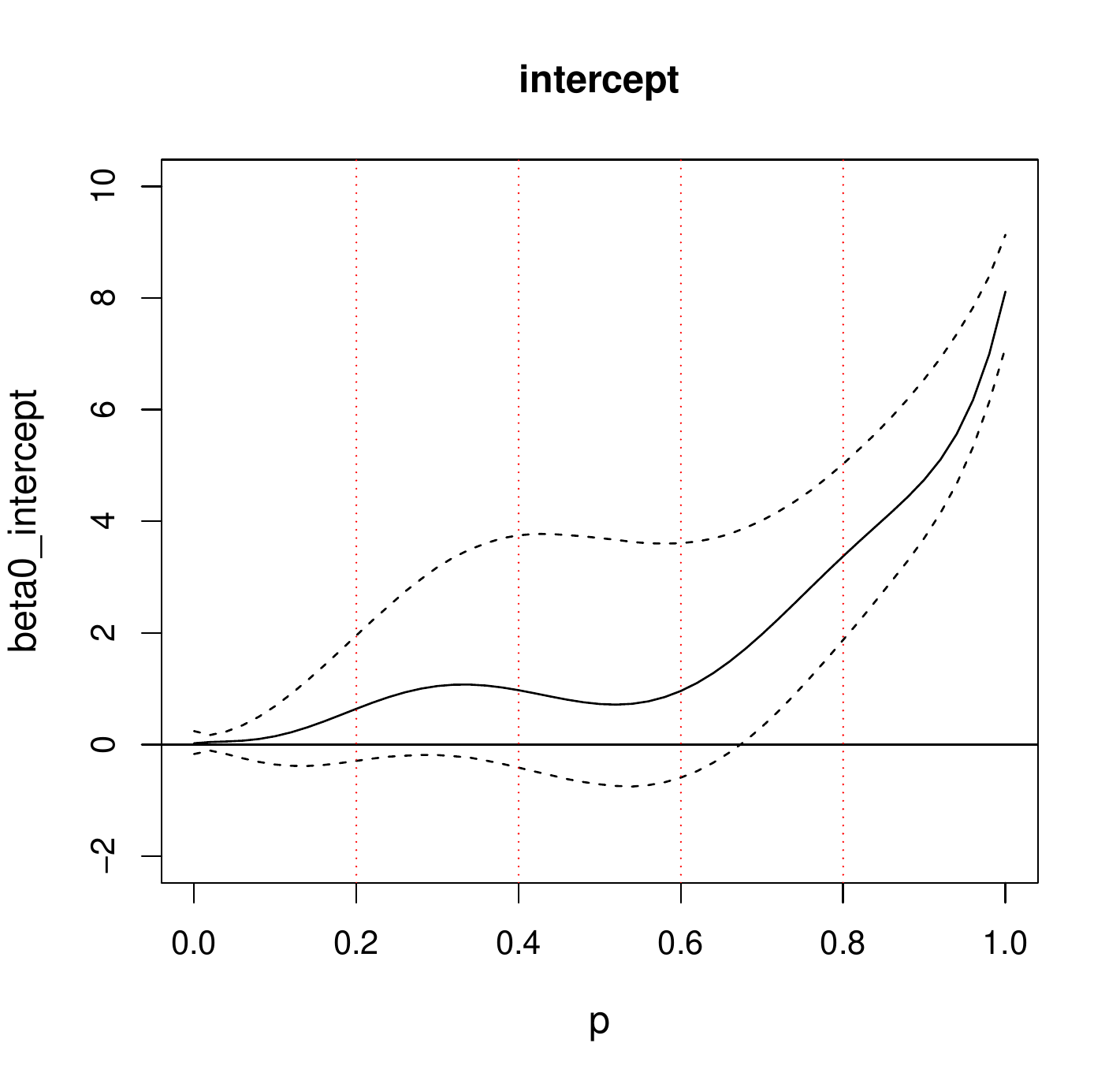} &
\includegraphics[width=.38\linewidth , height=.38\linewidth]{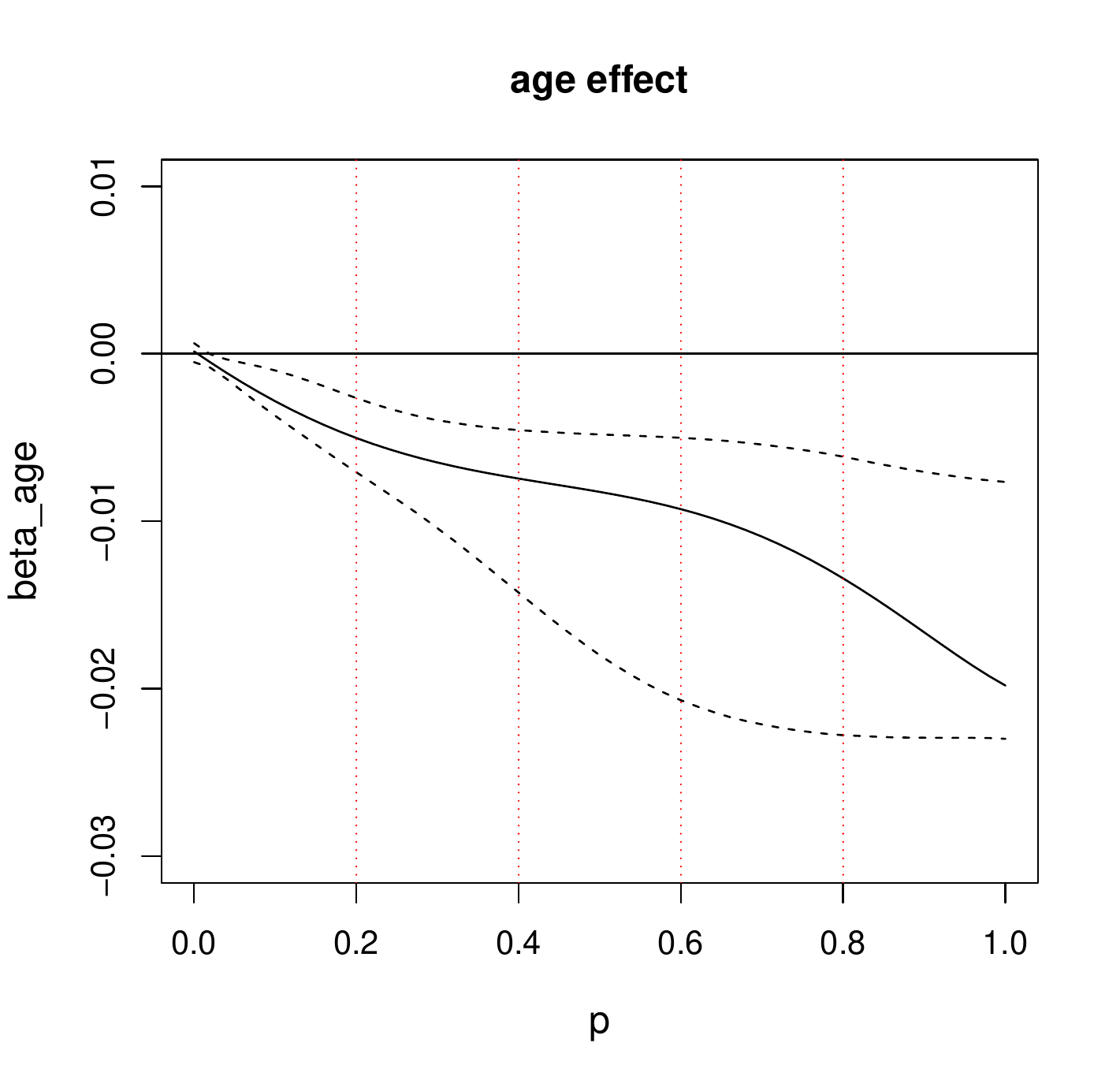}\\
\includegraphics[width=.38\linewidth , height=.38\linewidth]{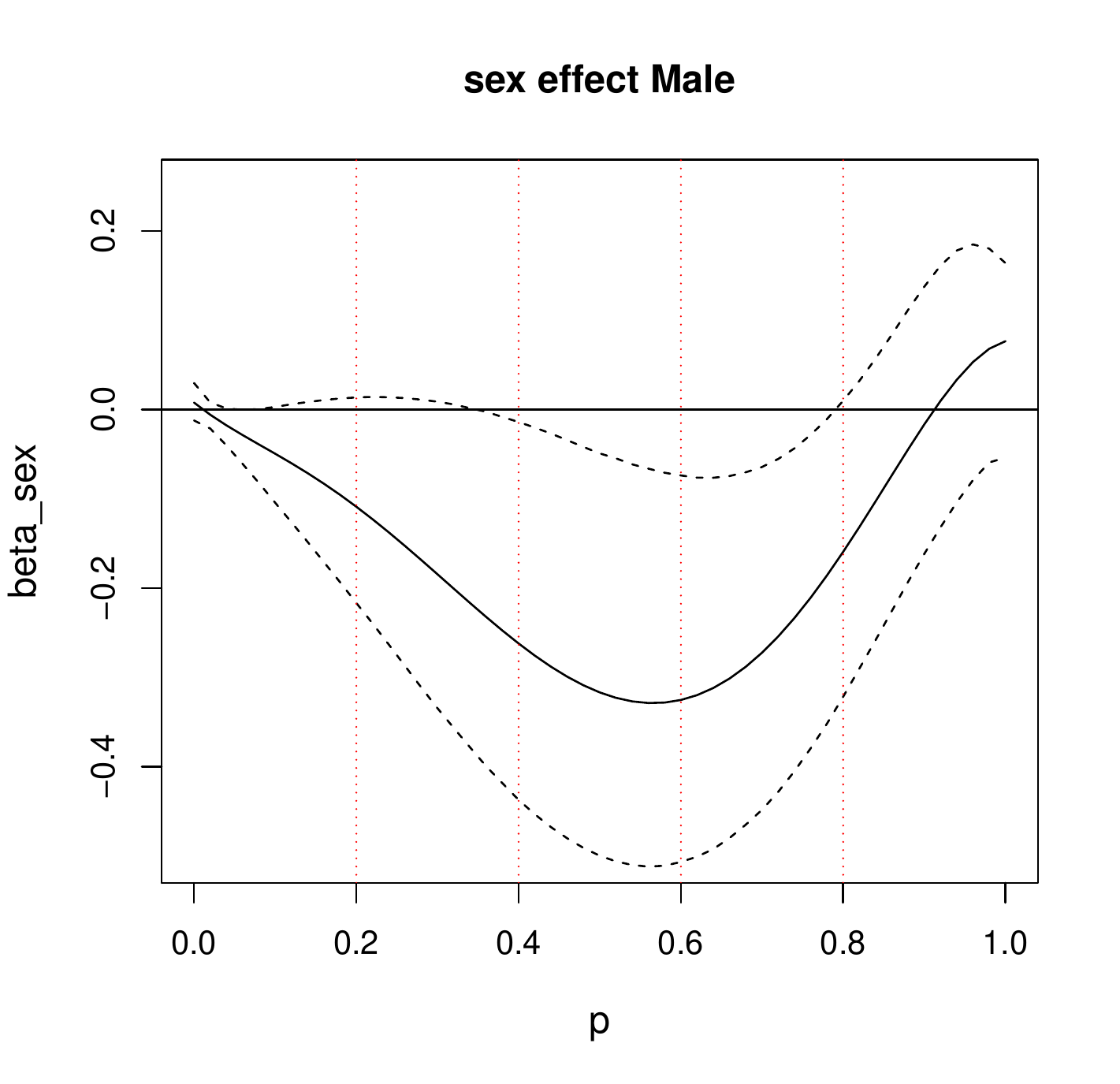} &
\includegraphics[width=.38\linewidth , height=.38\linewidth]{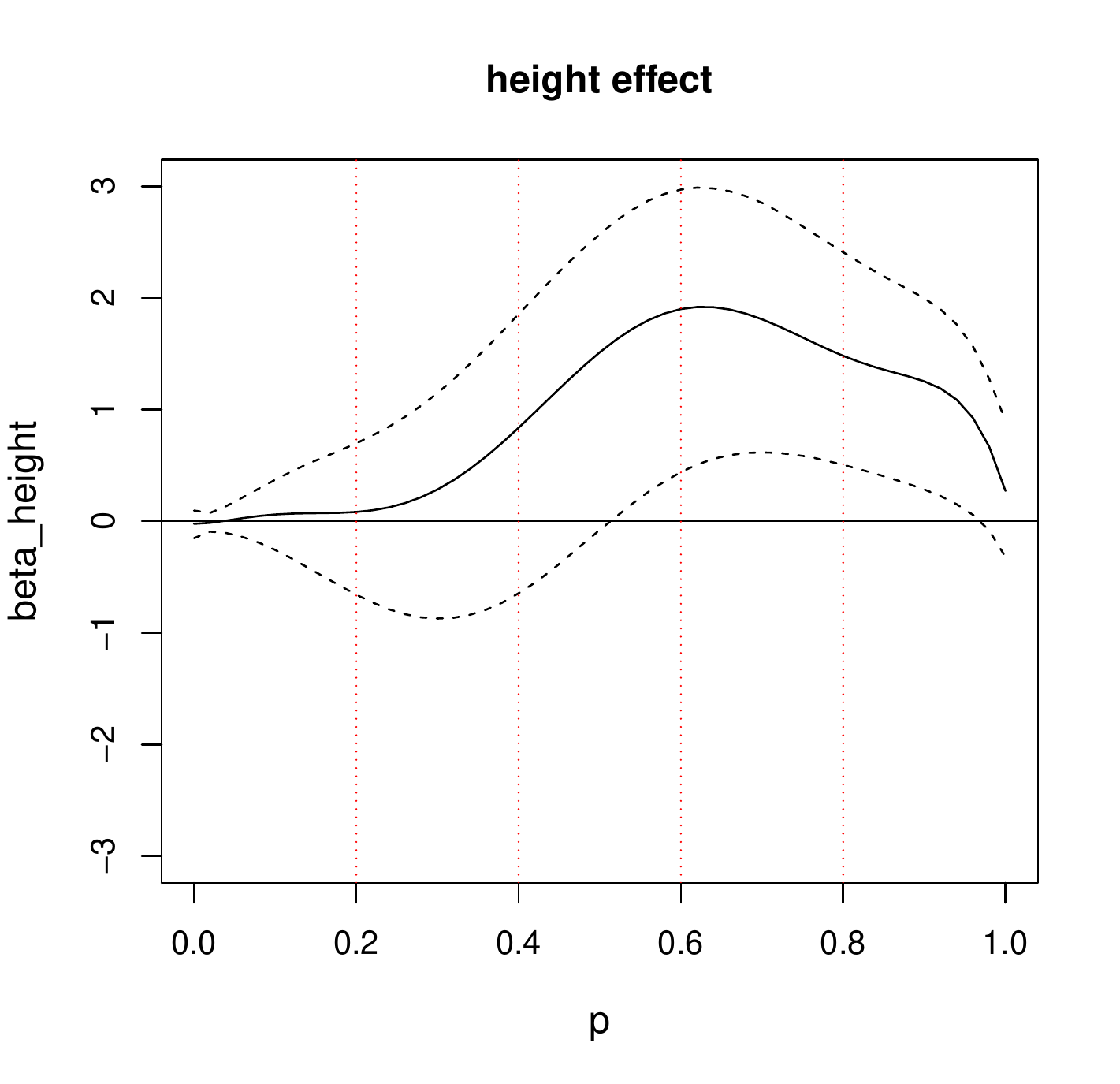}\\
\includegraphics[width=.38\linewidth , height=.38\linewidth]{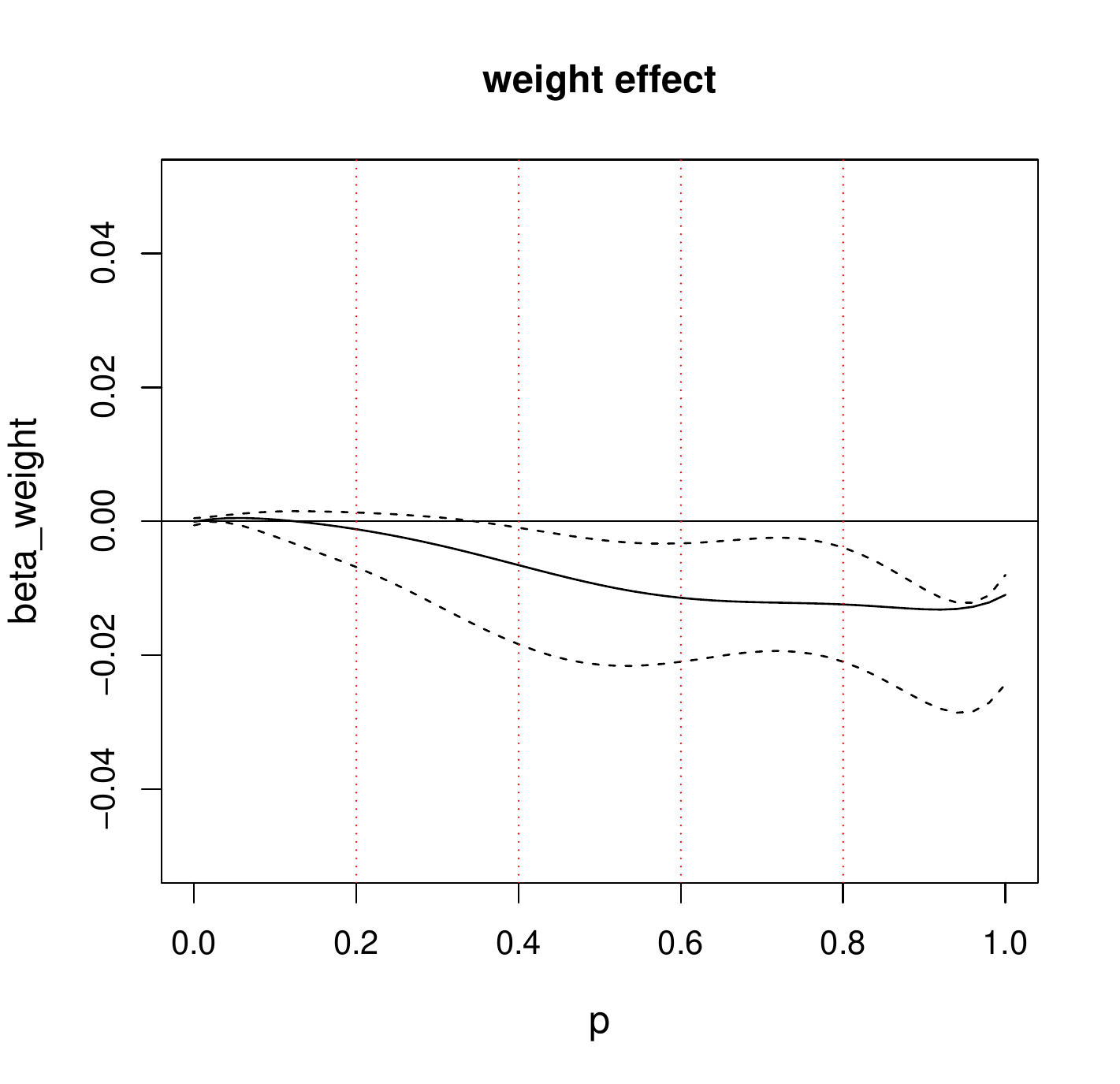} &  \\
\end{tabular}
\end{center}
\caption{Estimated quantile level effect of scalar (age, sex, height, weight) predictors on quantile function of physical activity in BLSA data using the quantile function-on-scalar regression model (13). Constrained BP estimates ($\beta_{age}(p)$ is decreasing) are shown in solid lines, along with their projection based $95\%$ confidence intervals (shown in dotted lines).}
\label{fig:fig2bl}
\end{figure}

We start with fitting unconstrained regression model (\ref{qfosr}) which is shown
in Figure S4. Several of the functional coefficients (e.g., age, gender etc.) are not necessarily non-decreasing. The age functional coefficient, in particular, appears to be decreasing, indicating an accelerated decrease in maximal levels of PA with increasing age \citep{varma2017re}. To further confirm this phenomenon, we test the null hypothesis $H_{0}: \beta_{age}(p) \hspace{2 mm} \textit{is decreasing} \hbox{ for all $p\in [0,1]$}$ using the proposed bootstrap test ($B=200$) in this article. The p-value of the test is calculated to be $0.12$, hence we fail to reject the null hypothesis that $\beta_{age}(p)$ is decreasing.

Therefore, next, we impose the monotonicity constraint that $\beta_{age}(p)$ is decreasing. We assume a common degree of smoothness (due to computational tractability) in the coefficient functions which is controlled by the order the Bernstein polynomial basis $N$. The common order of the Bernstein polynomial basis used to model all regression coefficient functions is chosen to be $7$ using the five-fold cross-validation method.

Figure \ref{fig:fig2bl} displays the estimated coefficient functions from the proposed Bernstein based constrained estimation approach along with their point-wise asymptotic $95\%$ confidence intervals constructed using the projection method. The estimated effect of age $\hat{\beta}_{age}(p)$ is found to be negative and decreasing over $p$ illustrating that physical activity not only decrease with age \citep{xiao2015quantifying,varma2017re}, but maximal levels of physical activity are decreasing with a faster rate. Based on the confidence intervals, the effect of age on PA is significant across all quantile levels.

The estimated effect of gender (Male) $\hat{\beta}_{sex}(p)$ indicates that males have higher maximal capacity of physical activity (for $p>0.8$) while females have higher activity levels in the range of $p$ between $0$ and $0.8$. In particular, females are shown to have significantly higher moderate levels of PA ($p \in (0.4,0.8)$) which is consistent with the findings of \cite{xiao2015quantifying}. For height, we see a significant positive effect $\hat{\beta}_{height}(p)$, especially, across a mid-range of $p$. The estimated effect  $\hat{\beta}_{weight}(p)$ is negative and appears to be decreasing, indicating an accelerated decrease of maximal levels of PA due to the increased weight.

\begin{figure}[H]
\begin{center}
\begin{tabular}{ll}
\includegraphics[width=.45\linewidth , height=.45\linewidth]{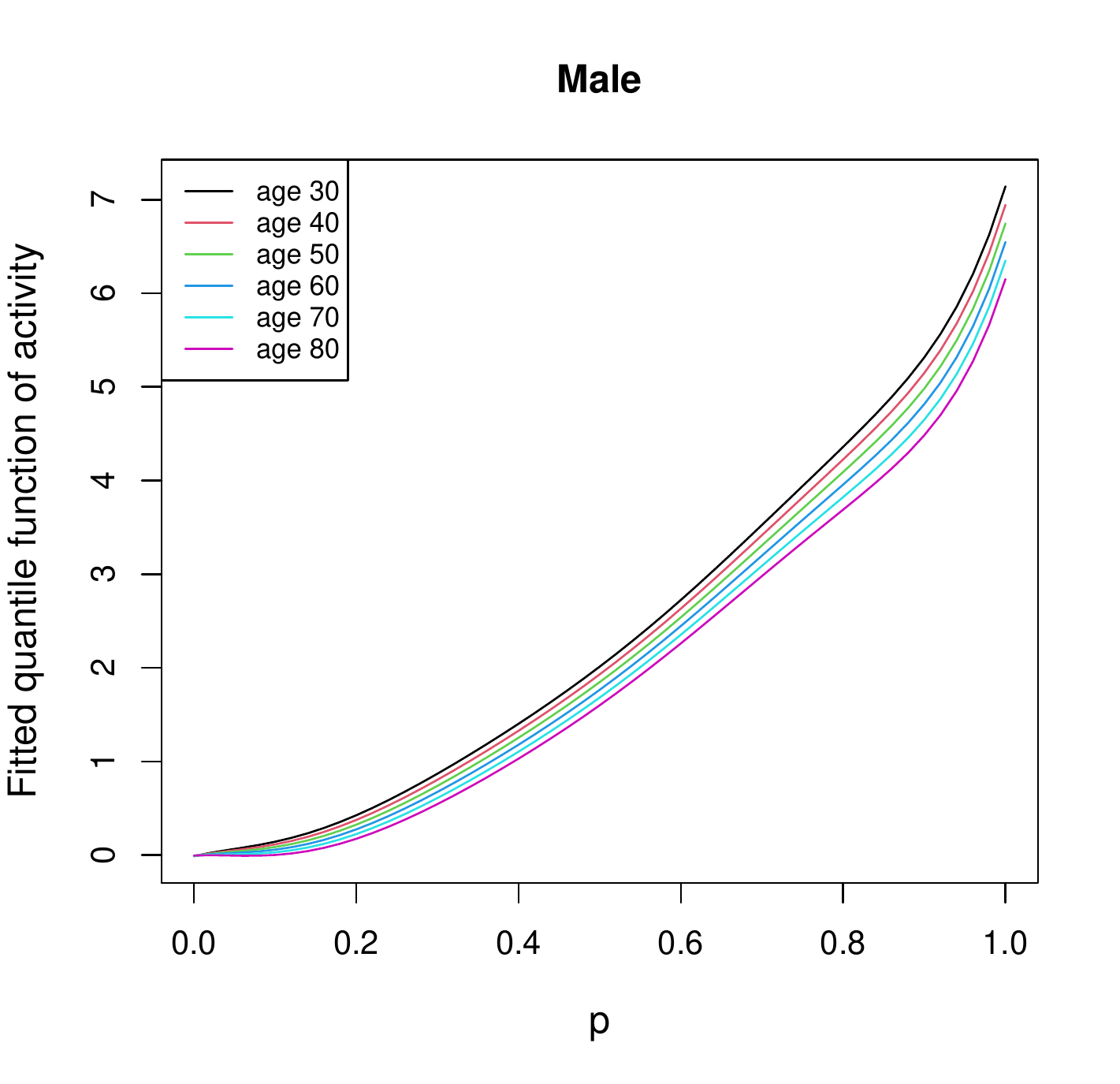} &
\includegraphics[width=.45\linewidth , height=.45\linewidth]{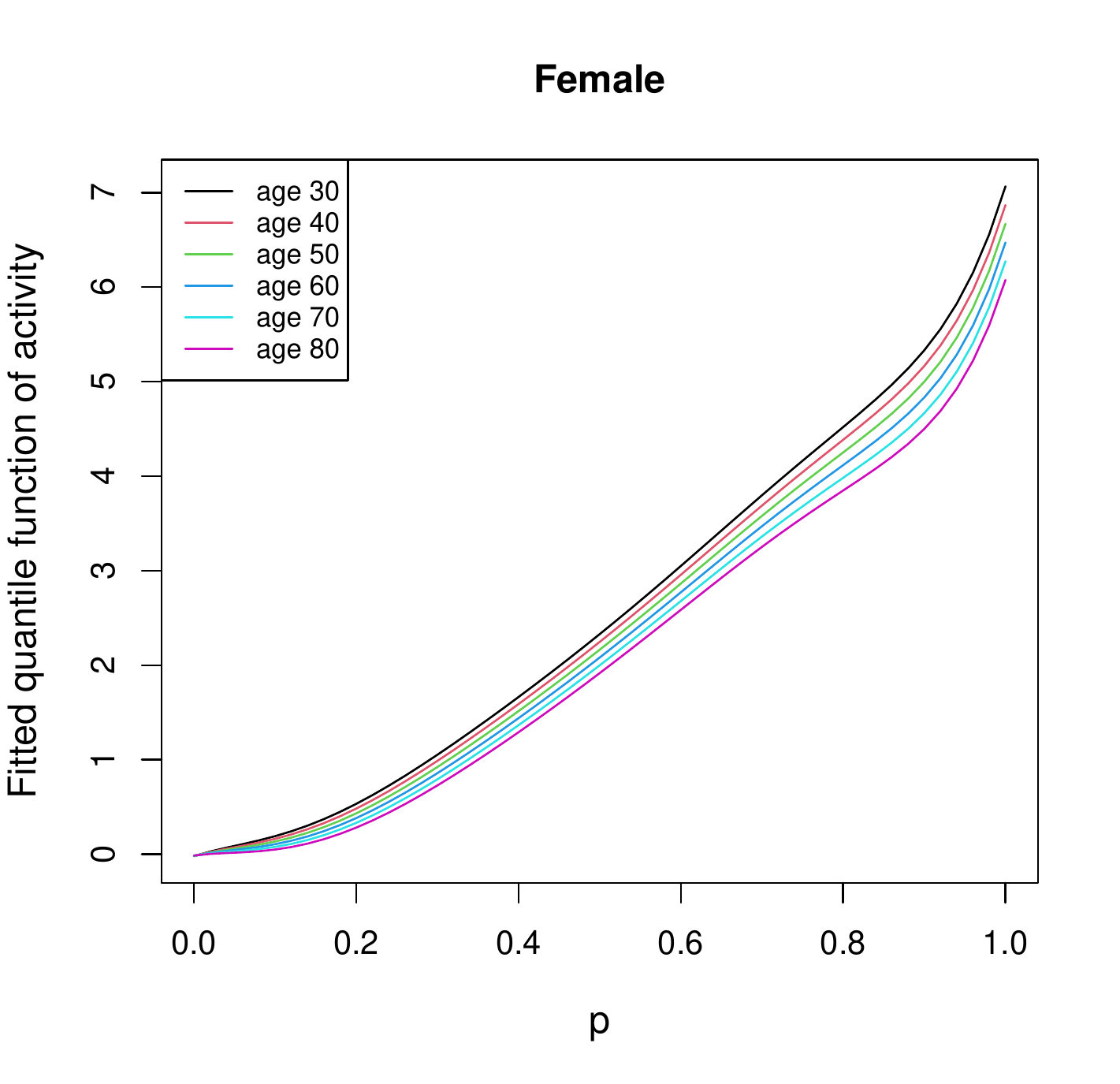}\\
\includegraphics[width=.45\linewidth , height=.45\linewidth]{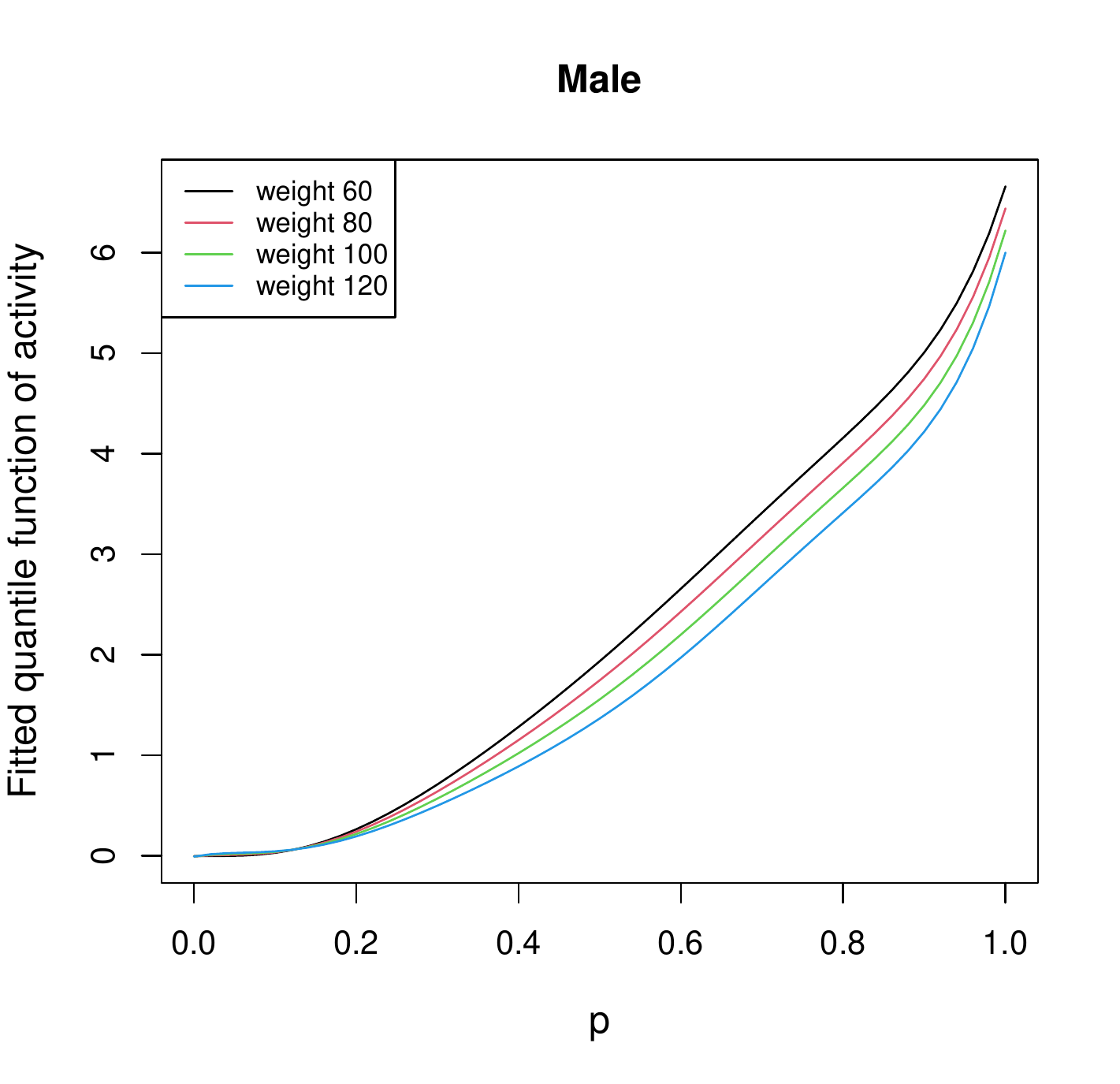} &
\includegraphics[width=.45\linewidth , height=.45\linewidth]{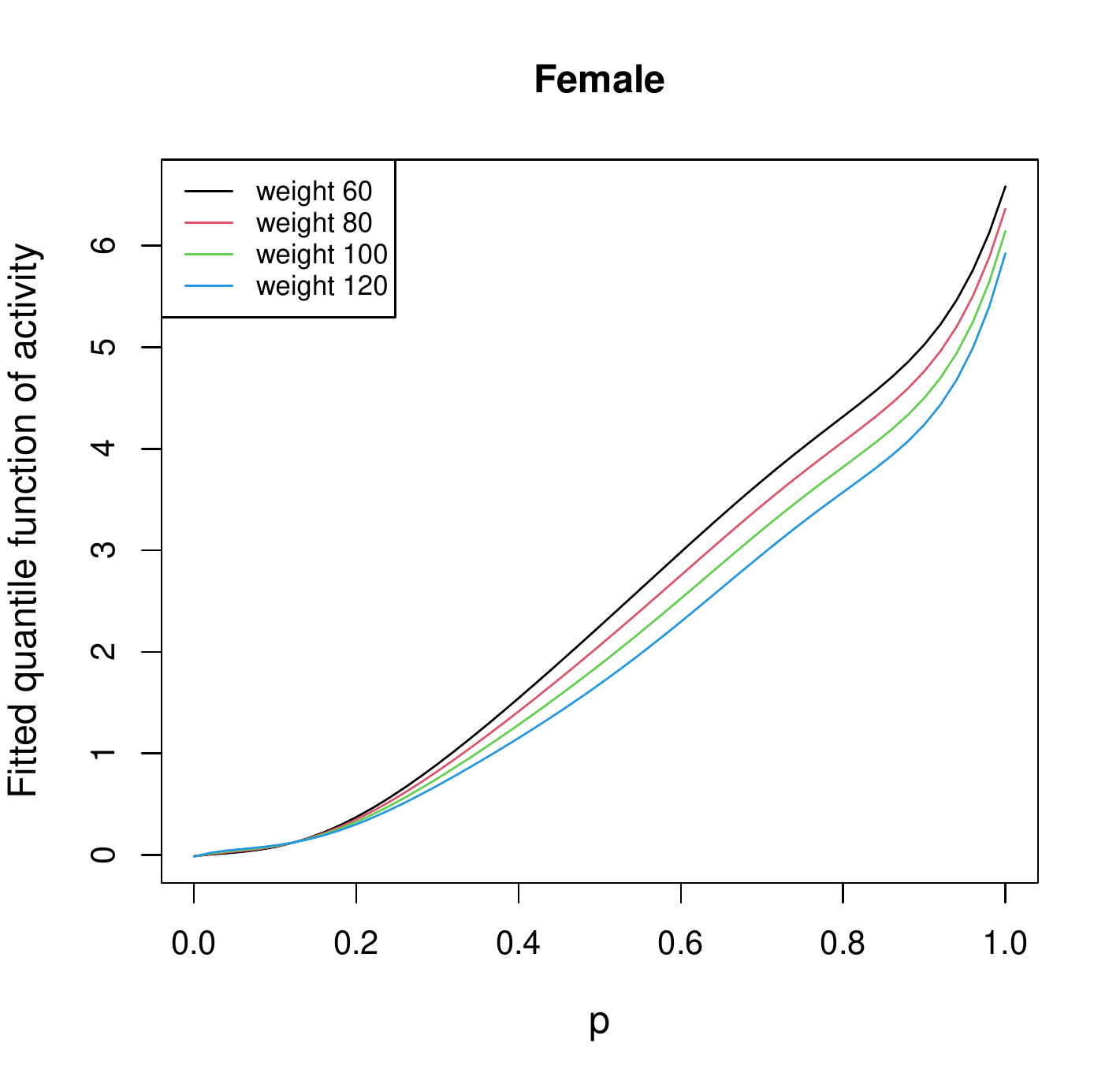}\\
\end{tabular}
\end{center}
\caption{Predicted quantile functions of physical activity for Male and Female for different ages at mean values of height and weight (top) and the predicted quantile function of physical activity for Male and Female for different weights at mean values of age and height (bottom).}
\label{fig:fig3bl}
\end{figure}

Fitted quantile functions stratified by gender are shown in Figure \ref{fig:fig3bl} across different ages and the centered  values of height and weight. Similarly, the predicted physical activity quantile function as function of weight is also displayed in Figure \ref{fig:fig3bl}. We notice the predicted quantile functions to be non decreasing and a clear separation among them with respect to age and weight. These results offer important scientific insights and a deeper understanding of dependence between subject-specific physical activity and demographic factors.


\hspace*{-8 mm}
\textit{Remark 5:}\\
Our analysis in this section illustrates that not all functional regression coefficients  (e.g., age) are non-decreasing in QFOSR. In Appendix D of Supplementary Material, we illustrate a method outlining a sufficient condition for the monotonicity of the quantile function which makes much weaker assumption compared to \cite{yang2020random}, allowing for a more flexible modeling of functional regression coefficients in QFOSR. We leave this as a future work to be explored more deeply.

\section{Discussion}
\label{disc}
We have developed a new estimation method for dealing with shape-constrained functional regression coefficients in common functional regression models. 
The estimation approach extends the one of \citep{wang2012shape}. It is shown that the key problem of shape restricted estimation is reduced to a linear inequality constrained least squares problem, where the constraint matrices are universal and do not depend on the value of the basis functions, the observed time-points, or the order of the splines unlike the B-spline based constrained estimation approaches \citep{ahkim2017shape}.

The proposed approach is computationally efficient and can be implemented with existing methods of quadratic programming. The constrained estimator is shown to be a projection of the unconstrained estimator, and consistency of the constrained estimator is established under the same regularity conditions as the unconstrained estimator. Projection-based point-wise asymptotic confidence intervals are developed for the constrained functional regression coefficients providing uncertainty quantification of the estimates. A residual bootstrap-based test is proposed that is based on the constrained estimation method. This further facilitates testing of various shape constraints in considered functional regression problems. Our empirical analysis illustrates that the proposed constrained estimation method can lead to reduced uncertainty of the functional coefficient in the restricted parameter space.

Applications shown on schizophrenia collaborative study and Baltimore Longitudinal Study of Aging illustrate the use of the proposed estimation method under prior constraints and offer important scientific insights into these problems. Although the estimation method is illustrated for functional data observed on a dense and regular grid, the method can be extended to more general scenarios where functional data are observed on irregular and sparse domains and covariates observed with measurement error. In such sparse setups, although, the individual number of observations $m_{i}$ can be small, and
$\bigcup_{i=1}^{n}\bigcup_{j=1}^{m_i}{t_{ij}}$ has to be dense in $\mathcal{T}$. Functional principal component analysis (FPCA) could be applied to de-noise the functional covariates and get their predicted trajectories at all time-points of interest \citep{ghosal2020variable}. 
The proposed estimation method for both functional and scalar response can then be used under the above scenarios. Examples include the schizophrenia collaborative study and the additional simulation set up under Scenario B of this article.

There are many research problems that remain to be explored based on this current work. One of the limitations of the current approach is the absence of a theoretically optimal choice for the order of the Bernstein polynomial basis $N$. Currently it is chosen in a data driven way and it has shown a satisfactory performance in our empirical results. We have considered three common functional regression models (SOFR, FOSR, FOFR) and shown the application of the proposed estimation method under shape constraints. The models illustrated in this article are building blocks of functional regression models \citep{Ramsay05functionaldata} and are limited in assuming linear effects of the predictors on the response, where as in many real world applications the effects might as well be nonlinear. Multiple extensions have been proposed in scalar-on-function \citep{reiss2010fast}, function-on-scalar and function-on-function regression models \citep{kim2016general} modeling dependence between scalar/functional response and scalar/functional covariates via unknown nonparametric functions. It would be of interest to explore the proposed Bernstein polynomial based estimation approach to extend to these models under prior shape constraints on the nonparametric functions.

Another interesting area of work would be
to extend the proposed estimation method beyond continuous response and allow scalar/functional responses coming from a general exponential family, e.g, binary data, count data etc. Using the proposed Bernstein polynomial based approach, in such cases would lead to optimizing a negative log-likelihood criterion under linear inequality constraints. One natural idea would be to use a second order Taylor approximation \citep{james2019penalized} of the log likelihood around the current estimate and solve iteratively a least square objective function under the linear inequality constraints. A Bayesian framework for generalized linear model could also be adapted to incorporate the prior restrictions \citep{ghosal2022bayesian}.
Extending the proposed shape constrained estimation method to such general classes of functional regression models would allow more diverse applications and remain areas for future research.

\section*{Software}
Software implementation via R \citep{Rsoft} and illustration of the proposed framework is available online with this article.

\section*{Supplementary Material}
Additional methodological illustrations and Supplementary Tables and Figures are available online with the Supplementary Material.

\section*{Acknowledgements}
The authors would like to thank the Associate Editor and two anonymous Reviewers for their constructive feedback which have led to an improved version of this manuscript.

\bibliographystyle{chicago}
\bibliography{Bibliography-MM-MC}
\clearpage
\newpage

\end{document}


\def\spacingset#1{\renewcommand{\baselinestretch}%
{#1}\small\normalsize} \spacingset{1}


\if0\blind
{
  \title{\bf Supplementary Material for Shape-Constrained Estimation in Functional Regression with Bernstein Polynomials}
 \author{Rahul Ghosal$^{1,\ast}$, Sujit Ghosh$^{2}$, Jacek Urbanek$^{3}$, \\Jennifer A. Schrack$^{4}$, Vadim Zipunnikov$^{5}$ \\
  $^{1}$ Department of Epidemiology and Biostatistics\\ University of South Carolina\\
   $^{2}$ Department of Statistics, North Carolina State University\\ 
 $^{3}$ Department of Medicine, Johns Hopkins University \\School of Medicine\\
 $^{4}$ Department of Epidemiology, Johns Hopkins Bloomberg \\School of Public Health\\
 $^{5}$ Department of Biostatistics, Johns Hopkins Bloomberg \\School of Public Health
  }
  \maketitle
} \fi

\if1\blind
{
  \bigskip
  \bigskip
  \bigskip
  \begin{center}
    {\LARGE\bf Title}
\end{center}
  \medskip
} \fi

\bigskip

\vfill

\newpage
\spacingset{1.5} 

\section{Appendix A: Proof of Theorem 1}
First, we will show that the constrained estimator $\hat{\beta}_{c}(t)$ is a consistent estimator of $\beta(t)$ as long as the unconstrained estimator $\hat{\beta}_{u}(t)$ is consistent. Second, conditions (H1)-(H5) will be used to verify the consistency of the unconstrained estimator using existing results for SOFR \citep{cardot1999spline,cardot2003spline}. Notice that for Bernstein polynomial basis approximation, the condition (H4) with just $q=v=0$ (i.e., $\mathcal{H}=C[0,1]$) would suffice.

The prior shape restrictions, e.g., positivity, monotonicity, convexity or their combinations on the regression coefficient function of the form $\beta(t) \in \mathcal{F}$ can be approximated by the linear constraints $\^A\bm\beta\geq \bm 0$ using the Bernstein polynomial basis expansion. Notice that although the matrix $\^A$ and the vector $\bm\beta$ varies with $n$, 
we suppress such notations for simplicity. The restricted parameter space is given by $ \bm\Theta_R=\{\bm\beta \in R^{K_n}: \^A\bm\beta\geq \bm 0  \}$, where again $K_n$ is allowed depend on the sample size $n$.
The basis coefficients are then estimated via constrained least squares with linear inequalities. The scalar-on-function regression model after basis expansion with Bernstein polynomials can be written (w.l.g by absorbing the intercept term) as follows 

 $$Y_i=\*W_i^T\bm\beta+\epsilon_i. \hspace{2mm} \label{scalar resp}$$
The unrestricted and restricted estimators are given by
\begin{equation}
    \hat{\bm\beta}_{ur}=\underset{\bm\beta \in R^{K_n} }{\text{argmin}}\hspace{2 mm}  \sum_{i=1}^{n}({Y_i}-\*W_i^T\bm\beta)^{2}, \hspace{4mm} \label{sofropt}
\end{equation}

\begin{equation}
    \hat{\bm\beta}_{r}=\underset{\bm\beta \in \bm\Theta_R }{\text{argmin}}\hspace{2 mm}  \sum_{i=1}^{n}({Y_i}-\*W_i^T\bm\beta)^{2}. \hspace{4mm} \label{sofropt2}
\end{equation}

We follow the proofs of \cite{meyer2008inference,freyberger2018inference} who showed that the constrained estimators are projection of the unconstrained estimators on the restricted space, particularly when the objective function is quadratic \citep{freyberger2018inference}. In our case, this can be shown by first noting that
$$\frac{1}{n}||\*Y-\^W\bm\beta||_2^{2}=\frac{1}{n}||\*Y-\^W\hat{\bm\beta}_{ur}||_2^{2}+\frac{1}{n}||\^W\hat{\bm\beta}_{ur}-\^W{\bm\beta}||_2^{2},$$
where $\*Y=(Y_1,Y_2,\ldots,Y_n)^T$ and $\^W=(\*W_1,\*W_2,\ldots,\*W_n)^T$. Hence, we have $\hat{\bm\beta}_{r}=\underset{\bm\beta \in \bm\Theta_R }{\text{argmin}}\hspace{2 mm} ||\bm\beta-\hat{\bm\beta}_{ur}||^{2}_{\hat{\*\Omega}},$ where
$\hat{\bm\Omega}=\frac{1}{n}\sum_{i=1}^{n}\*W_i\*W_i^T$ and $\bm\Omega=E(\hat{\bm\Omega})$ is non-singular. Let the Bernstein polynomial approximation of $\beta(t)$ be given by $\beta_N(t)= \sum_{k=0}^{N}\beta^0_{k}b_k(t,N)=\rho_{K_n}(t)^{'}\bm\beta^0$ for some $\bm\beta^0\in \bm\Theta_R$ \citep{wang2012shape,freyberger2018inference}, which uniformly converges to $\beta(t)$ (since $\mathcal{F}_{N}$ is nested in $\mathcal{F}$
and $\bigcup_{N=1}^{\infty}\mathcal{F}_{N}$ is dense in $\mathcal{F}$ with respect to the sup-norm).

Since the restricted estimator
$\hat{\bm\beta}_r$ is projection of the unconstrained basis coefficient $\hat{\bm\beta}_{ur}$,
we have:
$\Pr[||\hat{\bm\beta}_r -\bm\beta^0||_{\hat{\*\Omega}}>\epsilon] \leq \Pr[ ||\hat{\bm\beta}_r -\hat{\bm\beta}_{ur}||_{\hat{\*\Omega}} +  ||\hat{\bm\beta}_{ur} -\bm\beta^0||_{\hat{\*\Omega}}>\epsilon] \leq \Pr[||\hat{\bm\beta}_{ur} -\bm\beta^0||_{\hat{\*\Omega}} > \epsilon/2]$.
The reason is that $\hat{\bm\beta}_{r}$ is the projection of $\hat{\bm\beta}_{ur}$ and hence $||\hat{\bm\beta}_{r} -\hat{\bm\beta}_{ur}||_{\hat{\*\Omega}} \leq ||\hat{\bm\beta}_{ur} -\bm\beta^0||_{\hat{\*\Omega}}$. Let us also assume $\hat{\bm\Omega}=\frac{1}{n}\sum_{i=1}^{n}\*W_i\*W_i^T \rightarrow \*\Omega$, which is a standard regularity condition in the unconstrained case. The above inequalities directly show that $\hat{\bm\beta}_r$ is consistent, if the unrestricted estimator $\hat{\bm\beta}_{ur}$ is consistent. Conditions (H1)-(H5) guarantee consistency of the unconstrained estimator using existing results for SOFR \citep{cardot1999spline,cardot2003spline} and \cite{huang2004polynomial} (specifically, Theorem 3.1 in \cite{cardot2003spline}). Hence the restricted estimator is consistent under the same regularity conditions.

Moreover, if $\bm\beta^0\in \bm\Theta_R$ we have $$||\hat{\bm\beta}_{ur}-{\bm\beta}^0||^{2}_{\hat{\*\Omega}}=||\hat{\bm\beta}_{ur}-\hat{\bm\beta}_r||^{2}_{\hat{\*\Omega}}+||\hat{\bm\beta}_{r}-{\bm\beta}^0||^{2}_{\hat{\*\Omega}}.$$ Hence $||\hat{\bm\beta}_{r}-{\bm\beta}^0||^{2}_{\hat{\*\Omega}}\leq ||\hat{\bm\beta}_{ur}-{\bm\beta}^0||^{2}_{\hat{\*\Omega}},$ with equality iff $\hat{\bm\beta}_{ur}=\hat{\bm\beta}_{r}$. Hence the shape-restricted estimator of the basis coefficients has smaller squared
error loss than the unrestricted version, when the true regression function
satisfies the shape assumptions.

\section{Appendix B: Supplementary Results and Proofs for Functional Response Case}
We consider shape constrained estimation in the functional response regression model (11) of the paper, arising from function-on-scalar or function-on-function regression. We posit the following Theorem outlining consistency of the constrained estimator.

\begin{theorem}
Consider the functional response regression model (11). Suppose the following conditions hold.
\begin{enumerate}[label=(C\arabic*)]
\item The observed time-points  $S= \{t_{1},t_{2},\ldots, t_{m} \}\sim F_T$ on  $\mathcal{T}$; moreover they are independent of the response and covariate processes $\{Y_i(t),
X_{i}(t)\}$ ($i=1,2,\ldots,n$). The distribution $ F_T$ has a Lebesgue density $f_T(t)$ which is bounded away from $0$ and $\infty$ uniformly over $\mathcal{T}$.
\item  $\exists$ positive constants $M_1,M_2>0$ s.t the eigenvalues
$\lambda_0(t)\leq \lambda_1(t)$ of $W(t)= E\{H(t)H^T(t)\}$ ($H(t)=[1\hspace{1 mm} X(t)]$) satisfy $M_1\leq\lambda_0(t)\leq \lambda_1(t)\leq M_2$ uniformly over $\mathcal{T}$.
\item $\exists$ $M_3>0$ s.t $|X(t)|\leq M_3$ for $t\in \mathcal{T}$.
\item There is a constant $M_4$ s.t $E(\epsilon^2(t))\leq M_4 < \infty$.
\item  $\textit{lim sup}_{n} \hspace{4 mm}\frac{\underset{\ell=0,1}{max} N_{\ell}}{\underset{\ell=0,1}{min} N_{\ell}} < \infty$, where $N_{\ell}$ is there order of the Bernstein basis polynomial used to model $\beta_{\ell}(t)$.
\item The error process can be decomposed as $\epsilon(t)= V(t) + w_t,$ where $V(\cdot)$ is a mean zero stochastic process and 
and $w_t$ are measurement errors that are independent at different time
points and have mean zero and constant variance $\sigma^2$.
\item$\lim_{n}d(\beta_{\ell}(\cdot),\mathcal{F}_{N,\ell})=0$ ($\ell=0,1$) and $\lim_{n}N_nlog N_n/n=0$, where $d(\beta_{\ell}(\cdot),\mathcal{F}_{N,\ell})$ is defined as $d(\beta_{\ell}(\cdot),\mathcal{F}_{N,\ell})=\inf_{g\in \mathcal{F}_{N,\ell}} \sup_{t\in \mathcal{T}} |\beta_{\ell}(t)-g(t)|$ and $N_n=\underset{l\in\{0,1\}}{max}N_{\ell}$. 
\item  $\hat{\#\Sigma}$ is a consistent estimator of $\#\Sigma$ in the sense $||\hat{\#\Sigma}^{-1}- \#\Sigma^{-1}||_{2} = o_{p}(1)$ (spectral norm).
\end{enumerate} 
If the prior shape restriction on $\beta_{\ell}(t)$ is correct, i.e., the true coefficient function $\beta_{\ell}(t) \in \mathcal{F_{\ell}}$, then the constrained estimator $\hat{\beta}_{c,\ell}(t)$ is a consistent estimator of $\beta_{\ell}(t)$.
\label{thm:fofr const}
\end{theorem}

\textbf{Proof:}
We again show that the constrained estimator $\hat{\beta}_{c,\ell}(t)$ is a consistent estimator of $\beta_{\ell}(t)$ if the unconstrained estimator $\hat{\beta}_{u,\ell}(t)$ is consistent. Regularity conditions (C1)- (C8) provides the consistency of the unconstrained GLS estimator using Theorem 1 in \cite{huang2004polynomial}. Let $\bm\beta(t)= [\beta_0(t) \hspace{2 mm}\beta_1(t)]$. It is enough to show that $\hat{\bm\beta}_{c}(t)$ is a consistent estimator of $\bm\beta(t)$ if $\hat{\bm\beta}_{u}(t)$ is consistent.

The prior shape restrictions, e.g., positivity, monotonicity, convexity or their combinations on the regression coefficient functions of the form $\bm\beta(t) \in \mathcal{F}$ ($\mathcal{F}=\mathcal{F}_0\times \mathcal{F}_1$) can be reduced to linear constraints $\^A\bm\beta\geq \bm 0$ using the Bernstein polynomial basis expansion. Let the restricted parameter space is given by $ \bm\Theta_R=\{\bm\beta \in R^{K_n}: \^A\bm\beta\geq \bm 0  \}$. The basis coefficients are then estimated via linear inequality constrained generalized least squares. For the moment let us assume $\#\Sigma$ to be known, in practice we will use the consistent estimator $\hat{\#\Sigma}$ in its place. The functional response regression model after basis expansion with Bernstein polynomials and pre-whitening with covariance matrix can be written as,

\begin{equation*}
    \*Y_i^*= \^B_0^*\bm\beta_0+\^W_i^*\bm\beta_1 + \bm\epsilon_i^*. \label{func response}
\end{equation*}
Here $\*Y_i^*={\#\Sigma}^{-1/2}_{m\times m} \*Y_i$, $\^B_0^*={\#\Sigma}^{-1/2}_{m\times m} \^B_0$,  $\^W_i^*={\#\Sigma}^{-1/2}_{m\times m} \^W_i$ and $\bm\epsilon_i^*={\#\Sigma}^{-1/2}_{m\times m} \bm\epsilon_i$. Let  $\^Z_i^*=[\^B_0^* \hspace{2mm}\^W_i^*]$. The unrestricted and restricted estimators are given by
\begin{equation}
    \hat{\bm\beta}_{ur}=\underset{\bm\beta \in R^{K_n} }{\text{argmin}}\hspace{2 mm}  \sum_{i=1}^{n}||{\*Y_i^*}-\^Z_i^*\bm\beta ||_2^{2} \hspace{4mm} \label{fofur}
\end{equation}

\begin{equation*}
    \hat{\bm\beta}_{r}=\underset{\bm\beta \in \bm\Theta_R }{\text{argmin}}\hspace{2 mm}  \sum_{i=1}^{n}||{\*Y_i^*}-\^Z_i^*\bm\beta ||_2^{2} \hspace{4mm} \label{fofr}
\end{equation*}
We again show that the constrained estimators are projection of unconstrained estimates on the restricted space \citep{freyberger2018inference}. Let us denote
$\*Y^{*T}=(\*Y_1^*,\*Y_2^*,\ldots,\*Y_n^*)^T$ and $\^Z^*=[\^Z_1^{*T},\^Z_2^{*T},\ldots,\^Z_n^{*T}]^T$.
$$\frac{1}{n}||\*Y^*-\^Z^*\bm\beta||_2^{2}=\frac{1}{n}||\*Y^*-\^Z^*\hat{\bm\beta}_{ur}||_2^{2}+\frac{1}{n}||\^Z^*\hat{\bm\beta}_{ur}-\^Z^*{\bm\beta}||_2^{2}.$$
Hence we have $\hat{\bm\beta}_{r}=\underset{\bm\beta \in \bm\Theta_R }{\text{argmin}}\hspace{2 mm} ||\bm\beta-\hat{\bm\beta}_{ur}||^{2}_{\hat{\*\Omega}},$ where
$\hat{\bm\Omega}=\frac{1}{n}\sum_{i=1}^{n}\^Z_i^{*T}\^Z_i^{*}$ and $\bm\Omega=E(\hat{\bm\Omega})$ is non-singular.  Let the Bernstein polynomial approximation of $\bm\beta(t)$ be given by $\beta_N(t)=\rho_{K_n}(t)^{'}\bm\beta^0$ for some $\bm\beta^0\in \bm\Theta_R$ \citep{freyberger2018inference}. Since the restricted estimator
$\hat{\bm\beta}_r$ is projection of the unconstrained basis coefficient $\hat{\bm\beta}_{ur}$, arguing in the similar way as in the scalar response case,
we have :
$\Pr[||\hat{\bm\beta}_r -\bm\beta^0||_{\hat{\*\Omega}}>\epsilon] \leq \Pr[ ||\hat{\bm\beta}_r -\hat{\bm\beta}_{ur}||_{\hat{\*\Omega}} +  ||\hat{\bm\beta}_{ur} -\bm\beta^0||_{\hat{\*\Omega}}>\epsilon] \leq \Pr[||\hat{\bm\beta}_{ur} -\bm\beta^0||_{\hat{\*\Omega}} > \epsilon/2]$.
We again assume $\hat{\bm\Omega}=\frac{1}{n}\sum_{i=1}^{n}\^Z_i^{*T}\^Z_i^{*} \rightarrow \*\Omega$. The above inequality directly shows that $\hat{\bm\beta}_r$ is consistent if the unrestricted estimator $\hat{\bm\beta}_{ur}$ is consistent, and
the shape-restricted estimator is consistent under the same regularity conditions as the unconstrained estimator.

\section{Appendix C: }
\begin{algorithm}[ht]
\label{algo1}
\caption{Bootstrap algorithm for shape testing with functional response}
\begin{algorithmic}
\label{algo 2}
\STATE 1. Fit the unconstrained FOSR model (2) or the FLCM (4) using the Bernstein-polynomial representation in (11) and calculate the residuals $e_i(t_j)=Y_i(t_j)-\hat{Y}_i(t_j)$, for $i=1,2,\ldots,n$.
\STATE 2. Fit the constrained model corresponding to $H_{0}$ (the null) and estimate $\beta_0(t),\beta_1(t)$ using the following constrained minimization criteria (no pre-whitening), 
 $$(\hat{\bm\beta_0},\hat{\bm\beta_1})=\underset{\bm\beta_0,\bm\beta_0}{\text{argmin}}\hspace{2 mm}  \sum_{i=1}^{n}||{\*Y_i}-\^B_0\bm\beta_0-\^W_i\bm\beta_1 ||_2^{2} \hspace{4mm} \textit{s.t \hspace{ 4 mm} $\^A\bm\beta_1\geq b$}. $$
Denote the estimates $\hat{\beta}^{c}_{0}(t), \hat{\beta}^c_{1}(t)$.
\STATE 3. Compute test statistic $T$ (15) based on these null and full model fits, denote this as $T_{obs}$.
\STATE 4. Resample B sets of bootstrap residuals $\{e^*_{b,i}(t)\}_{i=1}^{n}$ from residuals $\{e_{i}(t)\}_{i=1}^{n}$ obtained in step 1.
\STATE 5. for $b = 1$ to $B$ 
\STATE 6. Generate functional response under the constrained null model for the FLCM (for FOSR $X_i(t)=X_i$) as
$$Y^*_{b,i}(t)=\hat{\beta}^{c}_{0}(t)+  X_i(t) \hat{\beta}^c_{1}(t) +e^*_{b,i}(t),$$

\STATE 7. Given the bootstrap data set $\{X_i(t),Y^*_{b,i}(t)\}_{i=1}^{n}$ fit the null and the full model to compute the test statistic $T^*_b$.
\STATE 8. end for
\STATE 9. Calculate the p-value of the test as $\hat{p}=\frac{\sum_{b=1}^{B} I(T^*_b \geq T_{obs})}{B}$.
\end{algorithmic}
\end{algorithm}

\section{Appendix D: Monotonicity of Quantile functions as outcomes in QFOSR}
Let $\mu(p)=\beta_0(p)+\sum_{j=1}^{J}x_j\beta_j(p), 0\leq p\leq 1$. We assume $0\leq x_j \leq 1, \forall j= 1,2,\ldots, J$, without loss of generality (otherwise achieved by linear transformation or by an activation function, e.g., $z_j=F_0(x_j)$ for a c.d.f $F_0$ on support of $x_j$). Now, $\mu^{\prime}(p)=\beta_0^{\prime}(p)+\sum_{j=1}^{J}x_j\beta_j^{\prime}(p)$. Let $\beta_0^{\prime}(p)=\sum_{k=1}^{m}\gamma_{0k}b_{k-1}(p,m-1)$ and $\beta_j^{\prime}(p)=\sum_{k=1}^{m}\gamma_{jk}b_{k-1}(p,m-1)$, where $b_{k-1}(p,m-1)={m-1 \choose k-1}p^{k-1}(1-p)^{m-k}$ and $\gamma_{jk}$'s are the differences of the original Bernstein basis coefficients $\beta_{jk}$ corresponding to $\beta_j(p)$'s. Thus,
$$\mu^{\prime}(p)=\sum_{k=1}^{m}\left(\gamma_{0k}+\sum_{j=1}^{J}x_j\gamma_{jk}\right)b_{k-1}(p,m-1).$$
Now, as $\gamma_{0k}+\sum_{j=1}^{J}x_j\gamma_{jk}$ is a linear function in $(x_1,x_2,\ldots,x_J) \in [0,1]^J$, by the well-known Bauer's principle the minimum is attained at the boundary points $B=\{(x_1,x_2,\ldots,x_J):x_j\in \{0,1\}\}$. Hence the required sufficient condition for monotonicity (non-decreasing) is $$\underset{(x_1,x_2,\ldots,x_j) \in B}{\text{argmin}}  (\gamma_{0k}+\sum_{j=1}^{J}x_j\gamma_{jk})\geq 0, \hspace{2mm} \forall k.$$

\subsection*{Examples}
\begin{itemize}
    \item $J=1$: The constraint reduces to, $\underset{x_1 \in \{0,1\}}{\text{argmin}}  (\gamma_{0k}+x_1\gamma_{1k})= min(\gamma_{0k},\gamma_{0k}+\gamma_{1k} )\geq 0, \logeq \gamma_{0k} \geq 0 \hspace{2 mm} \& \hspace{2 mm} \gamma_{0k}+\gamma_{1k} \geq 0 \hspace{ 2 mm} \forall k$. Note
    that this allows $\gamma_{1k}\leq 0$ for some $k$ and hence a non-decreasing $\beta_1(p)$ is not required.
    \item $J=2$: A sufficient condition is $\underset{(x_1,x_2) \in \{0,1\}^2}{\text{argmin}}  (\gamma_{0k}+x_1\gamma_{1k}+x_2\gamma_{2k}) \geq 0, \logeq \gamma_{0k} \geq 0 \hspace{2 mm}, \hspace{2 mm} \gamma_{0k}+\gamma_{1k} \geq 0 \hspace{ 2 mm}, \gamma_{0k}+\gamma_{2k} \geq 0, \hspace{ 2 mm} \gamma_{0k}+\gamma_{1k}+\gamma_{2k} \geq 0 \hspace{2 mm} \forall k$.
\end{itemize}
Notice that the sufficient condition above is linear in $\gamma_{jk}$'s and hence $\beta_{jk}$'s. In general, the condition can be expressed via the linear inequality constraint $\^A_{N}\bm\beta_{N}\geq \bm 0$ (for a suitable matrix $\^A_{N}$), the framework illustrated in the paper.

\section{Supplementary Simulation Scenario S1}
 We consider a data generating scenario similar as in Scenario B (dense data) of the paper, with now $\beta_1(t)=2.5$, a constant function. We consider estimation under the constraint $\beta_1(t)$ is decreasing, so that the true function lies on the boundary of the restricted parameter space. Sample size $n=100$ and Bernstein polynomial basis of order $N=5$ is considered for estimation and obtaining the projection based confidence intervals. The average estimated coverage from the asymptotic $95\%$ confidence intervals is found to be $0.946$, close to the nominal level. Figure S3 in this Supplementary Material displays the estimated coefficient function along with its point-wise confidence intervals for one of the Monte Carlo replication. The constant function is observed to be well captured by the proposed estimate and the confidence intervals.

\section{Supplementary Tables}

\begin{table}[H]
\centering
\begin{tabular}{|c|c|c|c|}
\hline
BP order (N) & Sample size (n=25) & Sample size (n=50) & Sample size (n=100)             \\ \hline
2              & 0.88 (0.09)     & 0.87 (0.06)      & 0.86 (0.04) \\ \hline
3            & 0.90 (0.11)     & 0.92 (0.07)      & 0.90 (0.05)             \\ \hline
\textbf{4}            & \textbf{0.91  (0.15)}   & \textbf{0.93 (0.09)}       & \textbf{0.96 (0.06)}              \\ \hline
5            & 0.92 (0.23)     & 0.95 (0.13)      & 0.96 (0.08)              \\ \hline
6            & 0.91 (0.41)      & 0.93 (0.19)      & 0.94 (0.11)            \\ \hline
pfr (unconstrained)            & 0.88 (0.14)      & 0.94 (0.10)       & 0.98 (0.07)              \\ \hline
\end{tabular}
\caption{Average estimated coverage (over the grid of 50 equispaced time-points in $\mathcal{T}=[0,1]$) of the projection-based $95\%$ point-wise confidence intervals, for various choices of the order of the Bernstein polynomial (BP) basis, scenario A. Average width of the confidence interval is given in the parenthesis. The average choice of $N$ from cross-validation for this scenario is highlighted in bold.}
\label{tab:my-tables1}
\end{table}

\begin{table}[ht]
\centering
\begin{tabular}{|c|c|c|c|}
\hline
BP order (N) & Sample size (n=25) & Sample size (n=50) & Sample size (n=100)             \\ \hline
3            & 0.76 (0.26)     & 0.70  (0.17)     & 0.61 (0.12)            \\ \hline
4            & 0.82 (0.29)      & 0.78 (0.20)      & 0.68 (0.13)             \\ \hline
\textbf{5}            &\textbf {0.91 (0.34)}     &\textbf{ 0.92 (0.23)}     & \textbf{0.92  (0.16)}           \\ \hline
6            & 0.91 (0.38)      & 0.93 (0.26)      & 0.93  (0.17)            \\ \hline
7            & 0.93 (0.43)      & 0.93 (0.29)      & 0.94  (0.19)           \\ \hline
8            & 0.94 (0.55)      & 0.94 (0.35)       & 0.94  (0.22)            \\ \hline
pffr (U)            & 0.63 (0.40)      & 0.69 (0.29)       & 0.67 (0.21)              \\ \hline
pffr-sandwich (U)            & 1.00 (1.42)      & 0.99 (1.00)       & 1.00 (0.71)              \\ \hline
\end{tabular}
\caption{Average estimated coverage (over the grid of 40 equispaced time-points in $\mathcal{T}=[0,1]$) of the projection-based $95\%$ point-wise confidence intervals, for various choices of the order of the Bernstein polynomial (BP) basis, scenario B. Average width of the confidence interval is given in the parenthesis. The average choice of $N$ from cross-validation for this scenario is highlighted in bold. Last two rows report the results from the default unconstrained (U) (\texttt{"pffr"}) method within \texttt{"refund"} and from using a sandwich estimator for approximate variances in \texttt{"pffr"}.}
\label{tab:my-tables2}
\end{table}

\begin{table}[ht]
\centering
\begin{tabular}{|c|c|c|c|}
\hline
Sample size (n) & Constrained method & Unconstrained method & P-value              \\ \hline
50              & 1.84 (1.37)     & 3.86 (2.93)      & $<2.2\times 10^{-16}$ \\ \hline
100             & 0.967 (0.65)     & 1.91 (1.38)      & $<2.2\times 10^{-16}$               \\ \hline
\end{tabular}
\caption{Average integrated mean square error ($\times$ 100) over 200 Monte-Carlo replications, scenario B, Sparse data. Standard errors of IMSE  ($\times$ 100) are reported in the parenthesis. P-values are obtained from two sample t-test.}
\label{tab:my-tableS3}
\end{table}

\section{Supplementary Figures}
\begin{figure}[H]
\begin{center}
\includegraphics[width=.9\linewidth , height=.9\linewidth]{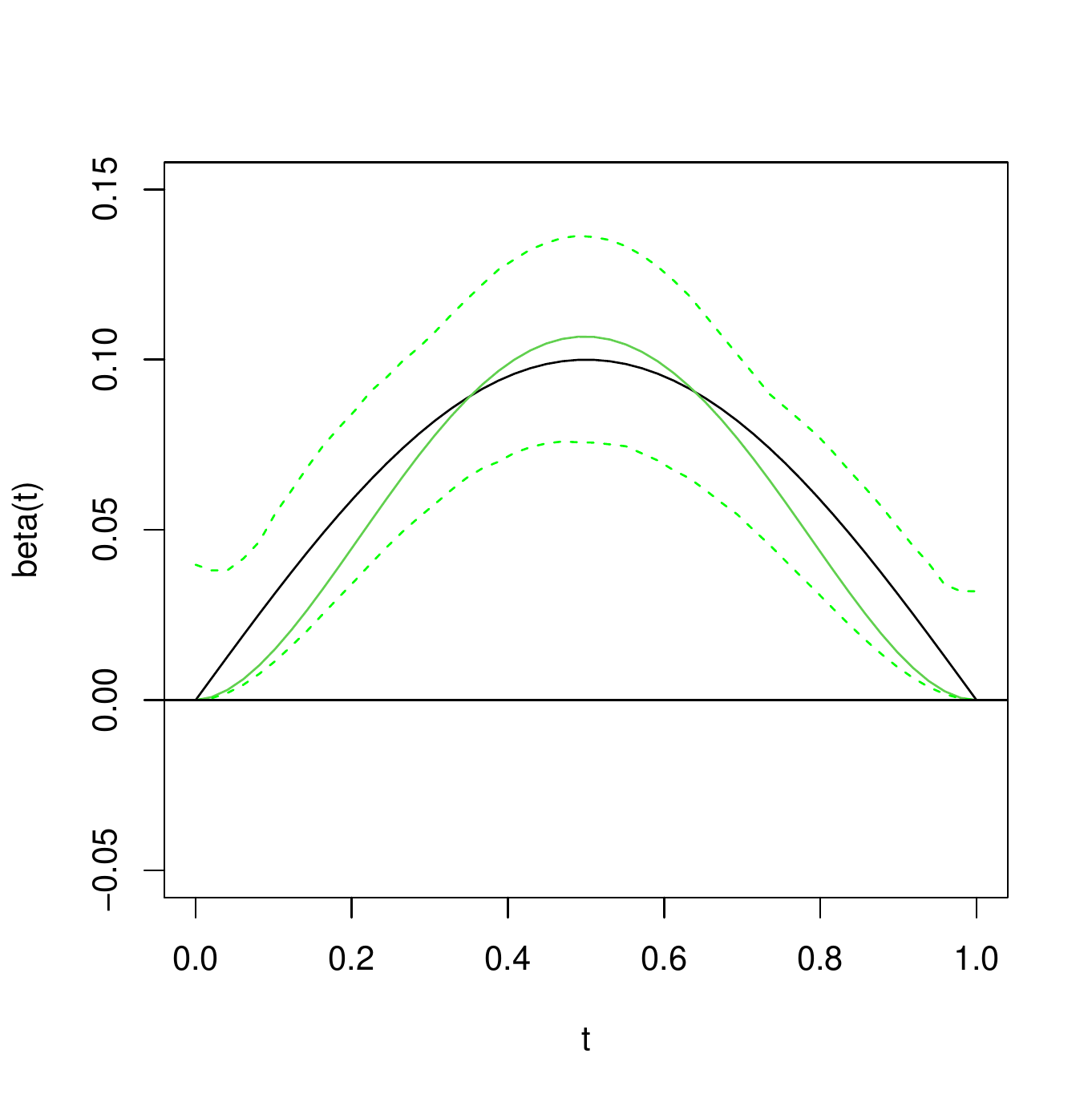}
\end{center}
\caption{Asymptotic $95\%$ point-wise confidence interval for $\beta(t)$ using the projection-based approach, scenario A, $n=100$. The true regression coefficient function is the solid black line, the constrained estimator is the solid green line, the dotted green lines show point-wise $95\%$ confidence intervals.}
\label{fig:figS3}
\end{figure}

\begin{figure}[H]
\begin{center}
\includegraphics[width=.9\linewidth , height=.9\linewidth]{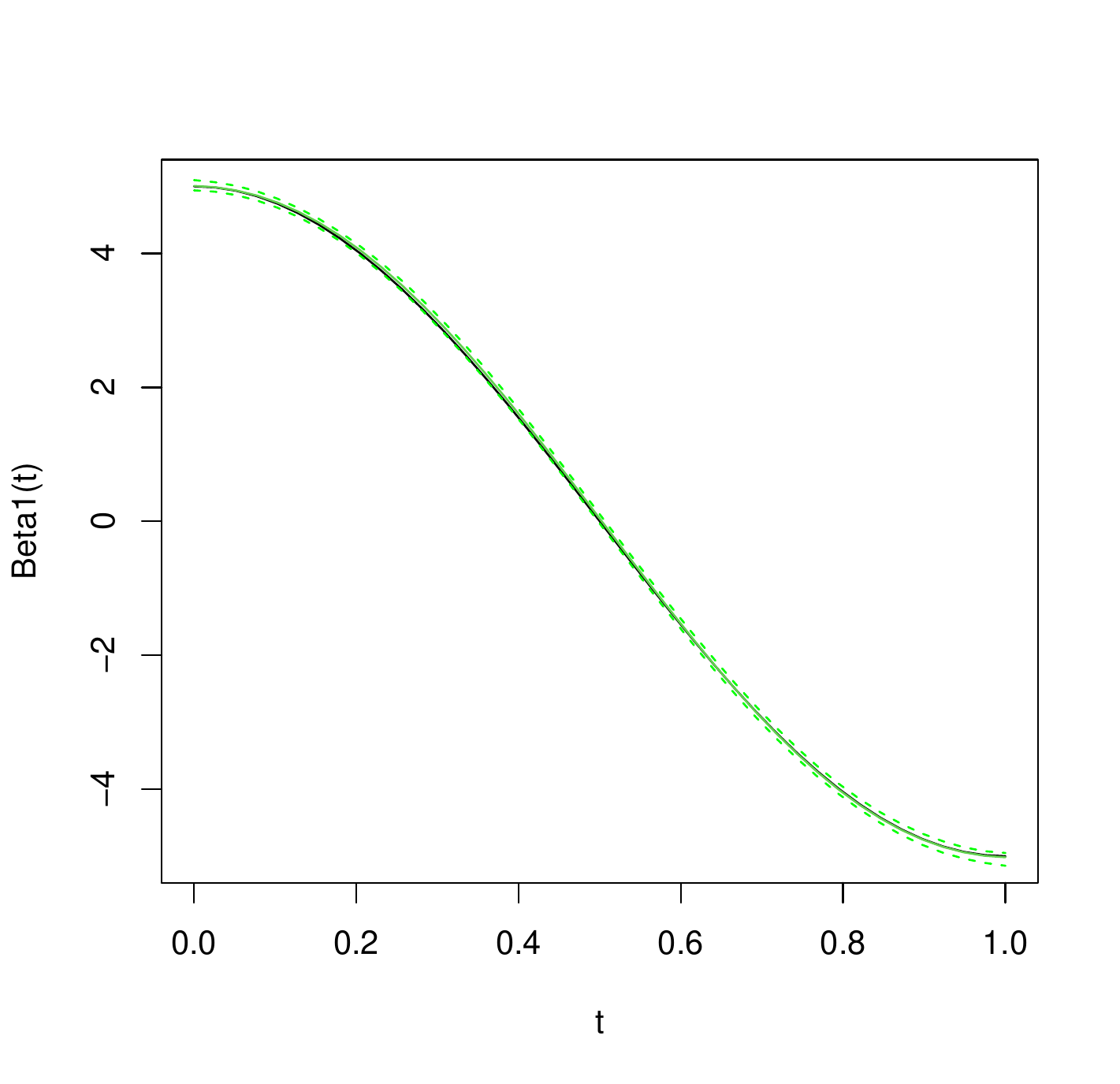}
\end{center}
\caption{Asymptotic $95\%$ point-wise confidence interval for $\beta_1(t)$ from the projection based approach, scenario B, $n=100$. The true regression coefficient function is the solid black line, the estimator is the solid green line, the dotted green lines show point-wise $95\%$ confidence intervals.}
\label{fig:figS3}
\end{figure}

\begin{figure}[H]
\begin{center}
\includegraphics[width=.9\linewidth , height=.9\linewidth]{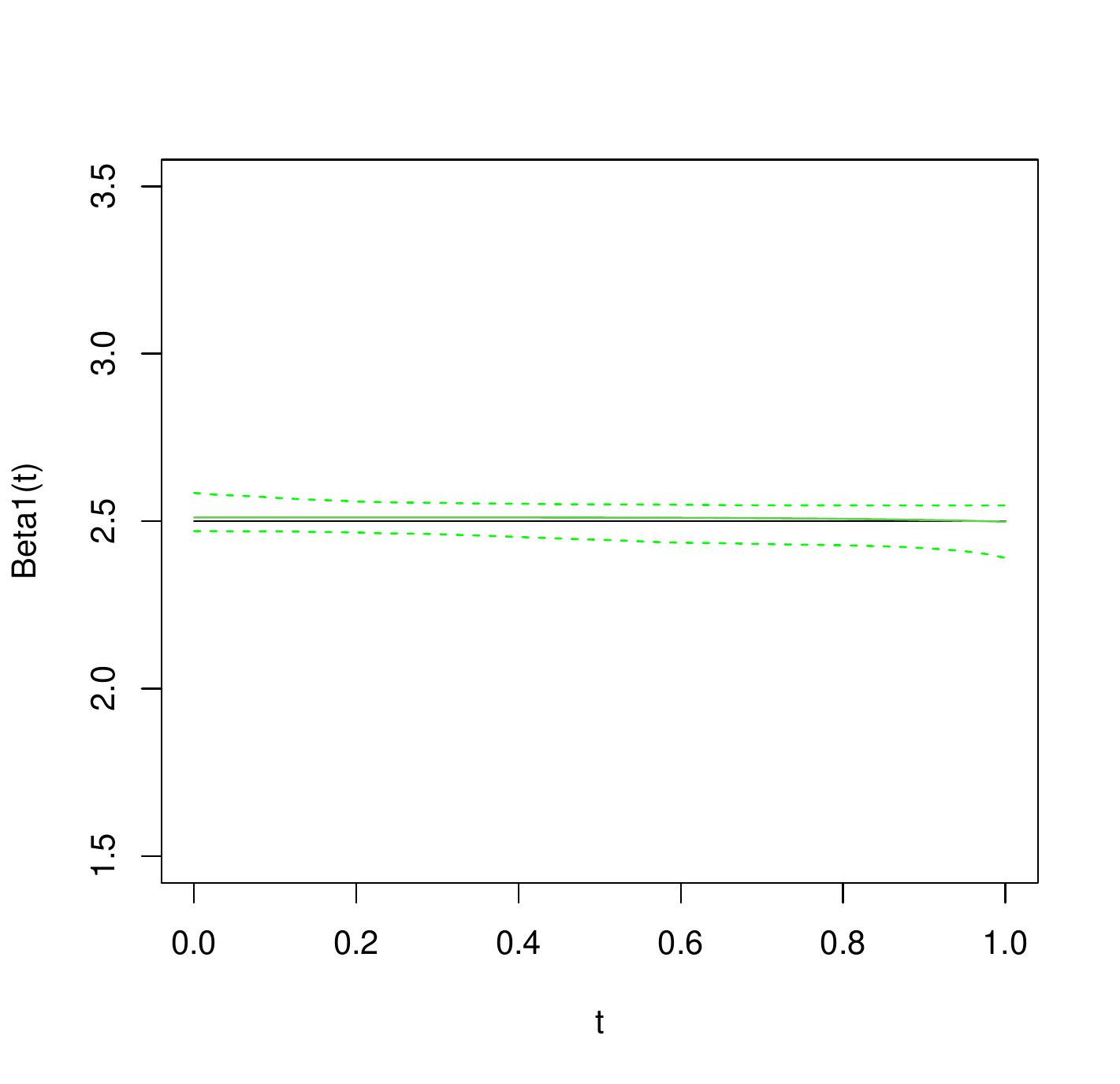}
\end{center}
\caption{Asymptotic $95\%$  point-wise confidence interval for $\beta_1(t)$ from the projection based approach, Supplementary scenario S1, $n=100$. The true coefficient function is shown via the solid black line and the estimate is shown in solid green, the dashed lines correspond to the point-wise $95\%$ confidence intervals.}
\label{fig:figS3}
\end{figure}

\begin{figure}[H]
\begin{center}
\begin{tabular}{ll}
\includegraphics[width=.38\linewidth , height=.38\linewidth]{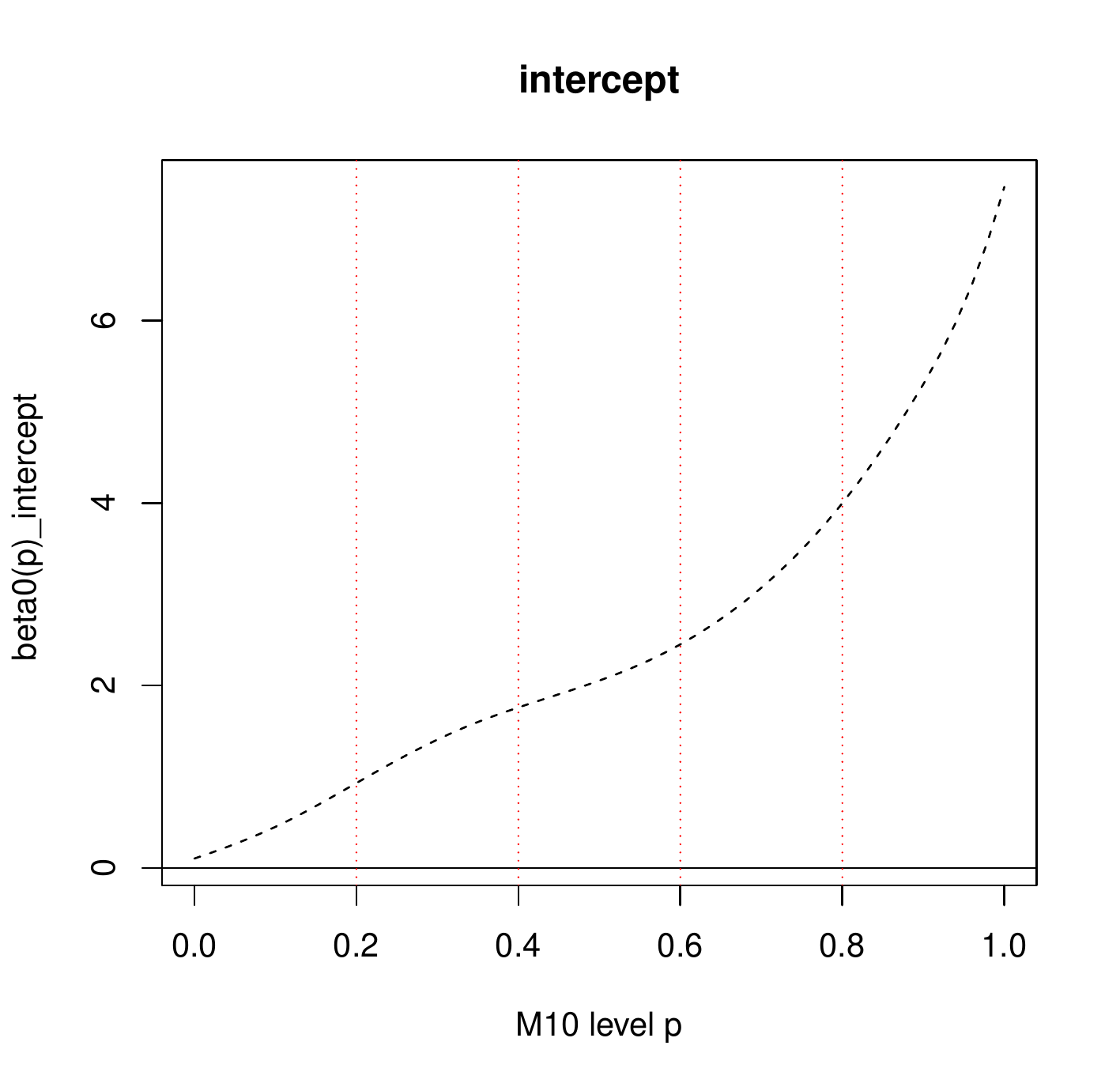} &
\includegraphics[width=.38\linewidth , height=.38\linewidth]{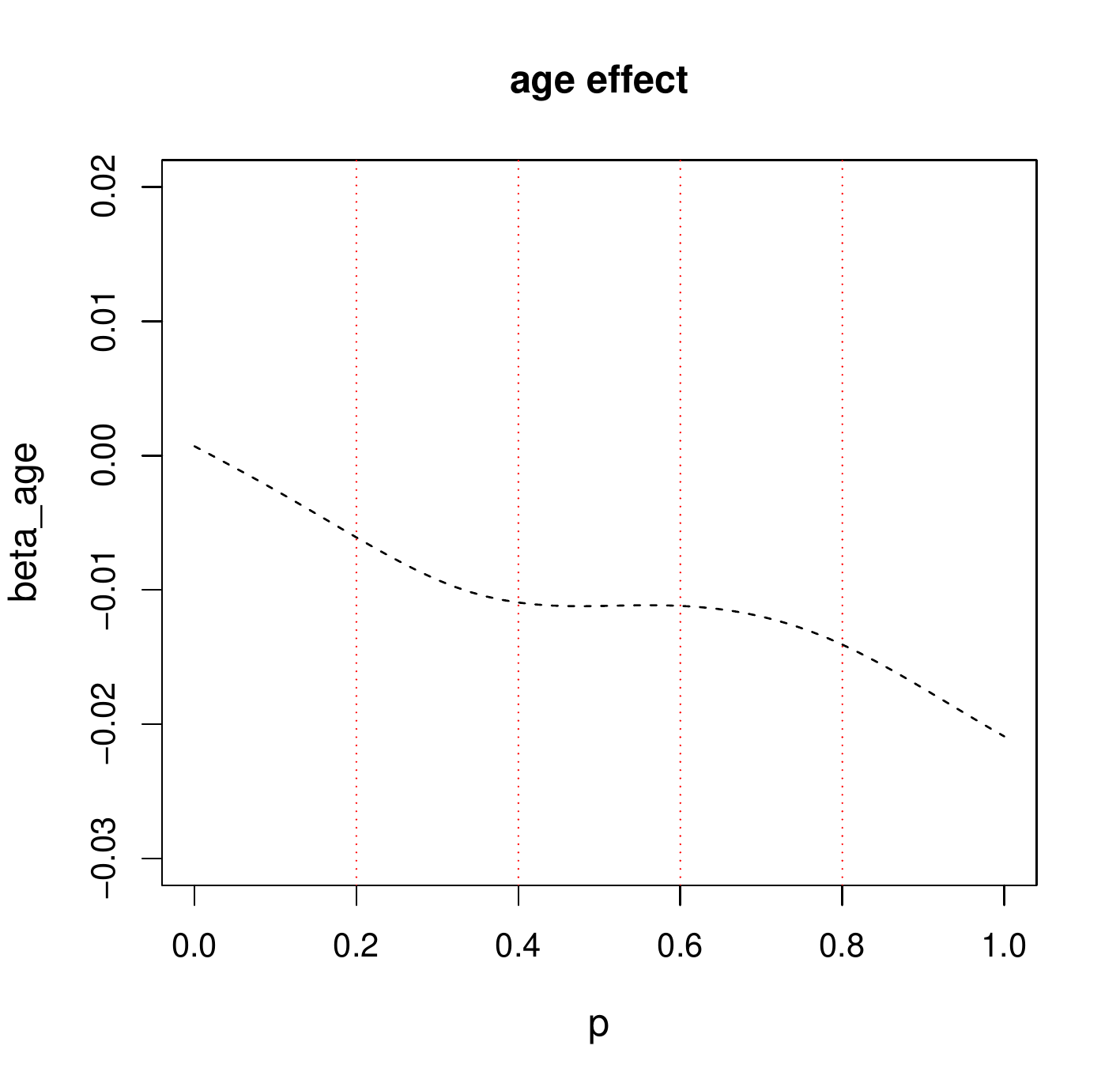}\\
\includegraphics[width=.38\linewidth , height=.38\linewidth]{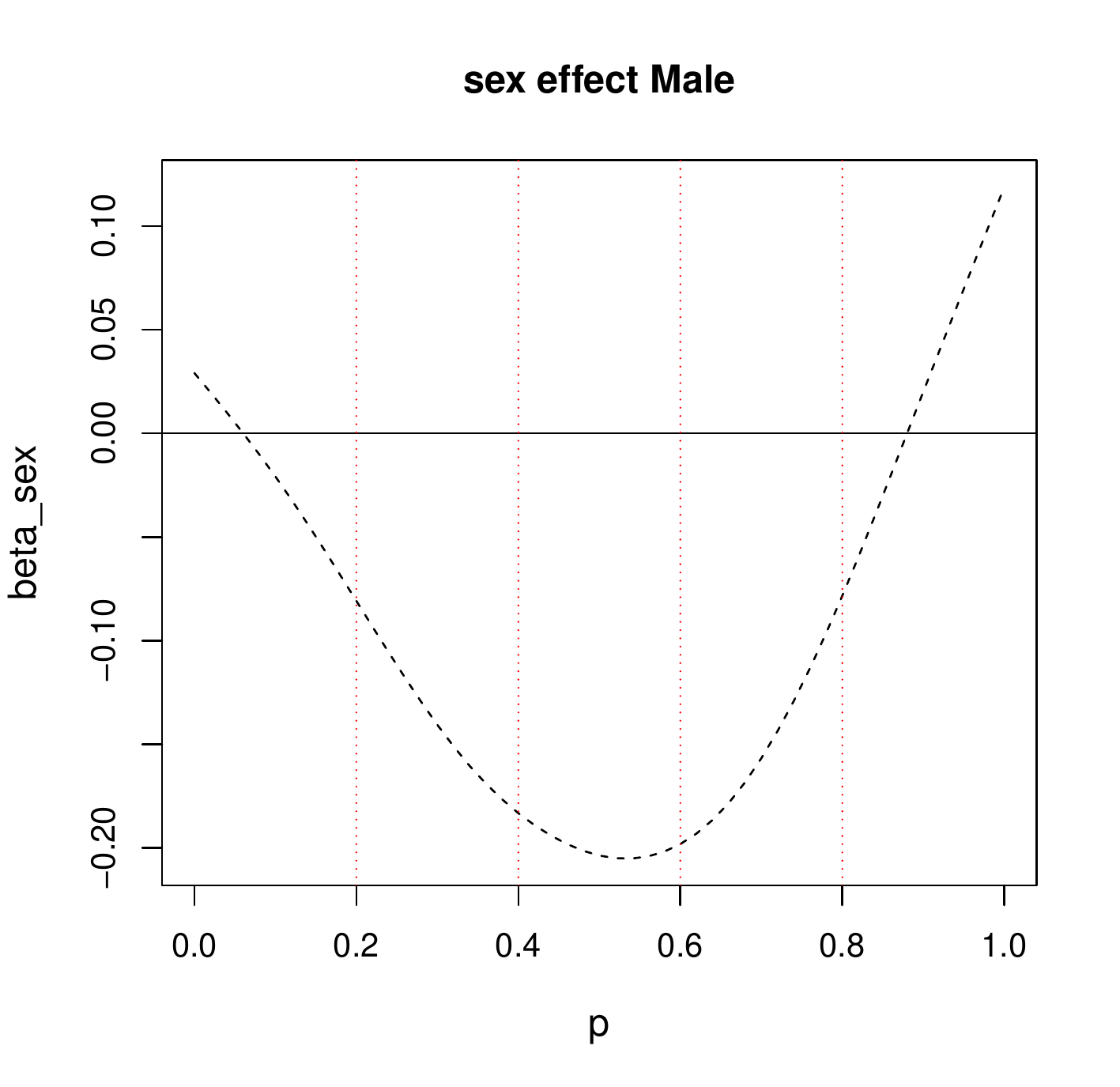} &
\includegraphics[width=.38\linewidth , height=.38\linewidth]{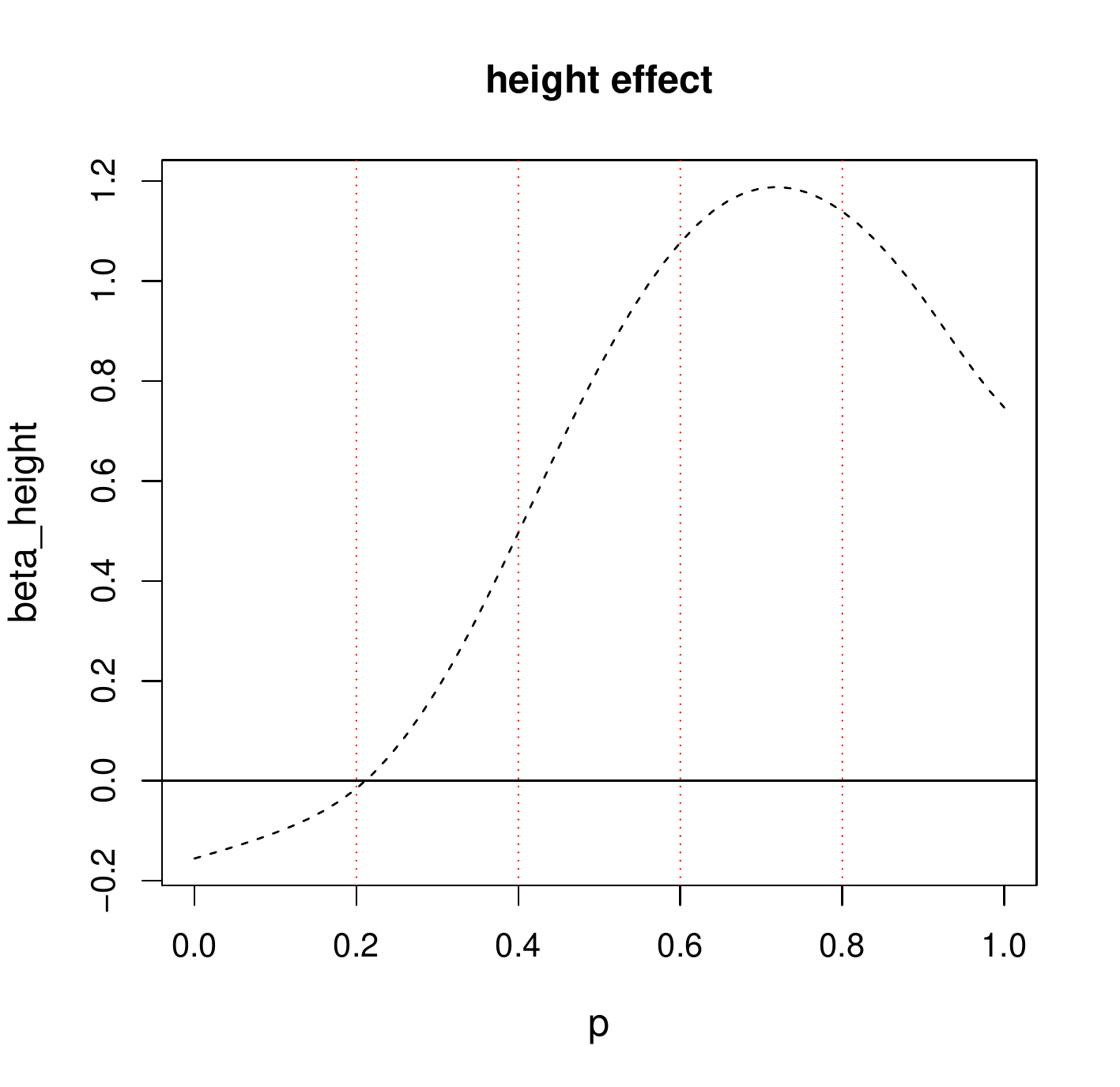}\\
\includegraphics[width=.38\linewidth , height=.38\linewidth]{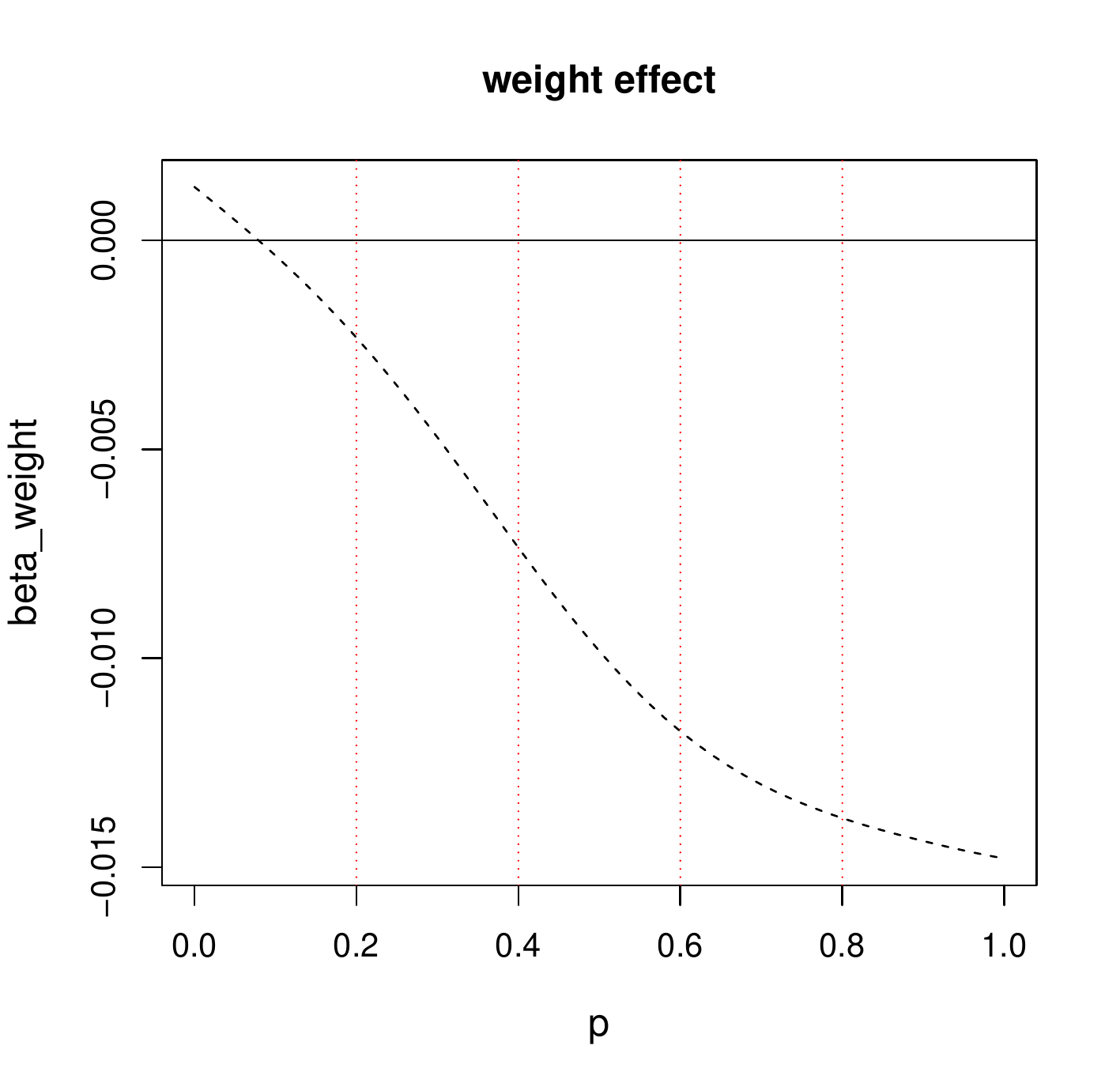} &  \\
\end{tabular}
\end{center}
\caption{Estimated quantile level effect of scalar (age, sex, height, weight) predictors on quantile function of physical activity in BLSA data using the quantile function-on-scalar regression model (13). Unconstrained penalized spline estimates obtained using the \texttt{"pffr"} function within the \texttt{refund} package are shown in dotted lines.}
\label{fig:figS3}
\end{figure}

\bibliographystyle{chicago}
\bibliography{Bibliography-MM-MC}